\documentclass[UTF8]{article}
\usepackage[utf8]{inputenc}
\usepackage{CJKutf8}
\usepackage[hyphens]{url}
\usepackage{xspace}
\usepackage{microtype}
\usepackage{graphicx}
\usepackage{subfigure}
\usepackage{booktabs} 
\usepackage{listings}
\usepackage{hyperref}

\usepackage[accepted]{icml2025}

\makeatletter
\def\Notice@String{
\textit{The second AI for MATH Workshop at the 42\textsuperscript{nd} International Conference on Machine Learning}, Vancouver, Canada. Copyright 2025 by the author(s).
}
\makeatother

\usepackage{amsmath}
\usepackage{amssymb}
\usepackage{mathtools}
\usepackage{amsthm}
\usepackage{enumitem}

\let\cite\citep

\usepackage[textsize=scriptsize]{todonotes}
\usepackage{xspace}

\makeatletter
\newcommand*\iftodonotes{\if@todonotes@disabled\expandafter\@secondoftwo\else\expandafter\@firstoftwo\fi}  
\makeatother

\definecolor{darkblack}{rgb}{0.0,0.0,0.5}
\definecolor{darkgreen}{rgb}{0.0, 0.42, 0.24}
\definecolor{lightgreen}{rgb}{0.52, 0.73, 0.4}
\definecolor{darkgray}{rgb}{0.4,0.4,0.4}
\definecolor{darkblue}{rgb}{0.0,0.0,0.5}
\definecolor{darkpurple}{rgb}{0.5,0.2,0.8}
\definecolor{lightpurple}{rgb}{0.8,0.5,1}

\usepackage{cleveref}
\crefname{figure}{Figure}{Figures}
\crefname{table}{Table}{Tables}
\crefname{appendix}{Appendix}{Apps.}
\crefname{section}{\S}{\S\S}
\crefformat{section}{\S#2#1#3} 
\crefname{equation}{Eq.}{Eqs.}
\crefname{algorithm}{Alg.}{Algs.}
\crefname{algocf}{Alg.}{Algs.}

\usepackage{wrapfig}
\usepackage{tabularx}
\usepackage{booktabs}
\usepackage{dsfont}
\usepackage{multirow}
\usepackage{array}

\usepackage{caption}
\usepackage{listings}
\usepackage{adjustbox}

\theoremstyle{plain}

\theoremstyle{definition}

\theoremstyle{remark}

\definecolor{framegray}{RGB}{170,170,170}
\icmltitlerunning{Lean Meets TCS}

\begin{document}
\begin{CJK}{UTF8}{gbsn}
\twocolumn[
\icmltitle{Lean Meets Theoretical Computer Science: \\Scalable Synthesis of Theorem Proving Challenges in Formal-Informal Pairs}


\icmlsetsymbol{equal}{*}
\icmlsetsymbol{supervisor}{$\dagger$}

\begin{icmlauthorlist}
\icmlauthor{Terry Jingchen Zhang}{equal,chab}
\icmlauthor{Wenyuan Jiang}{equal,yyy}
\icmlauthor{Rongchuan Liu}{equal,yyy}
\icmlauthor{Yisong Wang}{yyy}

\icmlauthor{Junran Yang}{comp}
\icmlauthor{Ning Wang}{yyy}
\icmlauthor{Nicole Ni}{Penn}
\icmlauthor{Yinya Huang}{yyy,hhh,supervisor}
\icmlauthor{Mrinmaya Sachan}{yyy,hhh,supervisor}
\end{icmlauthorlist}
\vspace{1cm}

\icmlaffiliation{chab}{D-CHAB, ETH Zurich, Zurich, Switzerland.}
\icmlaffiliation{yyy}{D-INFK, ETH Zurich, Zurich, Switzerland.}
\icmlaffiliation{hhh}{ETH AI Center, Zurich, Switzerland.}
\icmlaffiliation{Penn}{University of Pennsylvania, PA, USA.}
\icmlaffiliation{comp}{Independent Researcher.}
\icmlcorrespondingauthor{Yinya Huang}{yinya.huang@ethz.ch}

\icmlkeywords{Large Language Model, Formal Theorem Proving}
]
\printAffiliationsAndNotice{\icmlEqualContribution} 

\setlist[itemize]{noitemsep, topsep=0pt}

\begin{abstract}
Formal theorem proving (FTP) has emerged as a critical foundation for evaluating the reasoning capabilities of large language models, enabling automated verification of mathematical proofs at scale. However, progress has been constrained by limited datasets due to the high cost of manual curation and the scarcity of challenging problems with verified formal-informal correspondences. We propose leveraging theoretical computer science (TCS) as a scalable source of rigorous proof problems, where algorithmic definitions enable automated generation of arbitrarily many challenging theorem-proof pairs. We demonstrate this approach on two TCS domains: Busy Beaver problems, which involve proving bounds on Turing machine halting behavior, and Mixed Boolean Arithmetic problems, which combine logical and arithmetic reasoning. Our framework automatically synthesizes problems with parallel formal (Lean4) and informal (Markdown) specifications, creating a scalable pipeline for generating verified proof challenges. Evaluation on frontier models reveals substantial gaps in automated theorem proving: while DeepSeekProver-V2-671B achieves 57.5\% success on Busy Beaver problems, it manages only 12\% on Mixed Boolean Arithmetic problems. These results highlight the difficulty of long-form proof generation even for problems that are computationally easy to verify, demonstrating the value of TCS domains for advancing automated reasoning research.
\end{abstract}

\begin{figure}[ht]
    \centering    
    \includegraphics[width=1\linewidth]{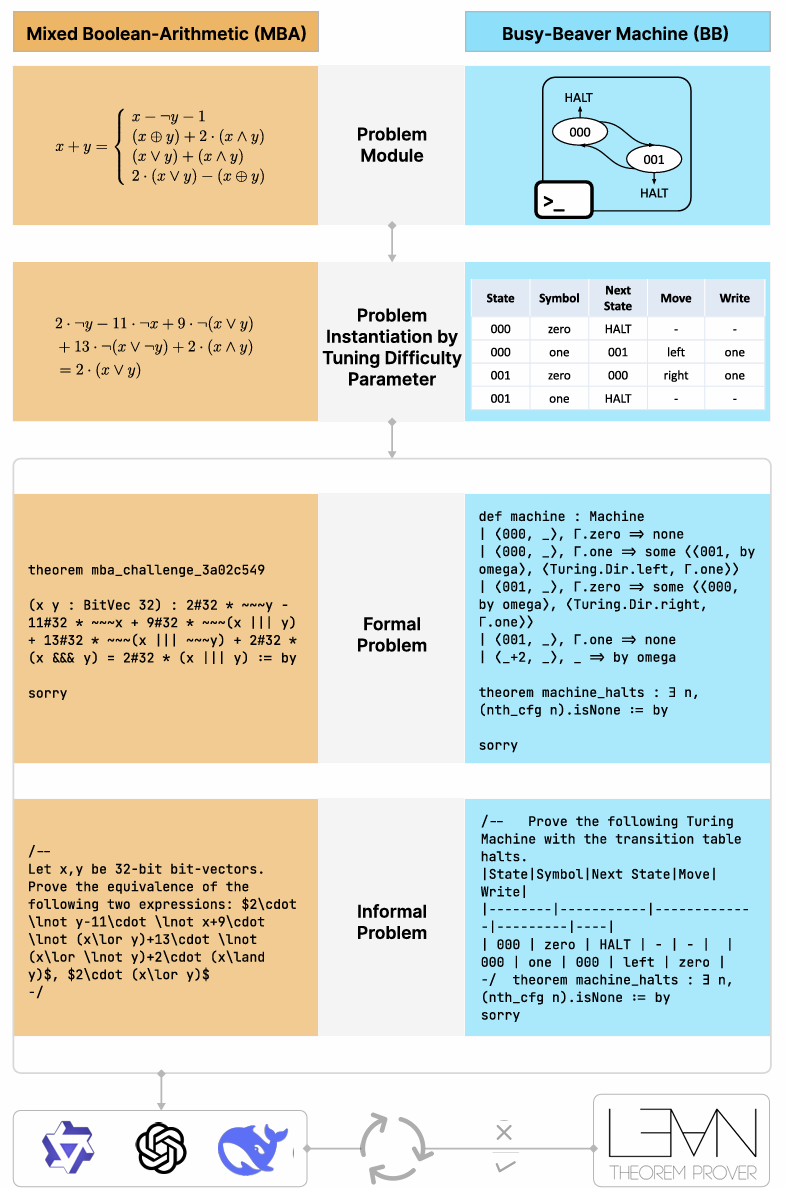}
    \caption{Overview of our synthesis framework: A problem is instantiated using expert-defined templates to generate theorem-proving challenges, with the formal description in Lean4 and the natural language description in Markdown.}
    \label{fig:SynthesisPipeline}
\end{figure}
\section{Introduction}

Capability evaluation of Large Language Models (LLMs) reasoning is increasingly shadowed by rapid saturation of benchmarks that rely on only final answers such as MATH~\cite{MATH} and GPQA~\cite{GPQA}, which cannot fully validate logical correctness of intermediate steps. Formal Theorem Proving (FTP)~\cite{FTPReview} emerges as a new frontier to automate stepwise validation of logical derivation by harnessing the power of formal theorem provers such as Lean~\cite{Lean4} and Isabelle~\cite{Isabelle}. However, current FTP datasets such as MiniF2F~\cite{miniF2F}, ProofNet~\cite{ProofNet}, and PutnamBench~\cite{PutnamBench} remain limited in scope and diversity due to laborious manual curation of public data prone to contamination~\cite{Contam1,Contam2} that inflate benchmark scores~\cite{ContaminationDetection} and obscure true model capabilities.

We propose Theoretical Computer Science (TCS) as a powerful and underutilized testbed for structured reasoning evaluation. We explore TCS problems such as Busy Beaver (BB) and Mixed Boolean Arithmetic (MBA) problems, which are uniquely suited for benchmark development as they admit precise algorithmic modeling, are easily verifiable through proof assistants, and allow for automatic synthesis with tunable complexity. This enables us to construct infinitely scalable benchmark problems that are fully automated, contamination-resistant through exponential parameter variation, and designed to probe generalization and abstraction by targeting specific reasoning strategies with controllable complexity gradients.

We introduce a new framework that leverages TCS problem modules to systematically evaluate the limits of formal reasoning in LLMs. Our work makes three major contributions:

\paragraph{A scalable and contamination-resistant benchmarking framework.} We present a synthesis pipeline that programmatically generates theorem-proving challenges as rigorously aligned formal-informal pairs, eliminating risks of hallucination by grounding all problems in rigorous semantics validated by proof assistants. This enables fine-grained control over problem difficulty and supports experiments on reasoning generalization across complexity gradients, opening a new research paradigm for studying how models scale their reasoning as problem structures become deeper and more abstract.

\paragraph{Systematic evaluation revealing abstraction failure modes in frontier models.} We evaluate several advanced provers, including Kimina-Prover-Preview~\cite{KiminaProver}, Goedel-Prover-SFT~\cite{Goedelprover}, and DeepSeek-Prover-v2-671B across our TCS challenges. While these models perform well on standard tasks, they fail dramatically on our benchmarks (e.g., 12\% on MBA and 0\% on several modules), exposing critical limitations where current benchmarks overestimate generalization ability.

\paragraph{A step-level analysis protocol revealing reasoning decomposition bottlenecks.} We introduce controlled step-level evaluation using an Out-of-Distribution (OOD) lemma library to probe abstract thinking capabilities. Our analysis shows that models like OpenAI-o4-mini achieve 98.88\% accuracy at individual reasoning steps, yet fail at assembling those steps into complete proofs. This isolates the transition from local reasoning competence to global proof construction as a fundamental bottleneck, revealing failure modes that existing benchmarks cannot detect.

\section{Related Work}

\paragraph{Synthetic Benchmark Against Contamination} Rising concerns about whether benchmarks reflect genuine reasoning capability of LLMs~\cite{Contam1,Contam2} have motivated various approaches to detect and mitigate data contamination~\cite{ContaminationDetection}. LiveBench~\cite{LiveBench} refreshes benchmark questions every 6 months, while MixEval~\cite{MixEval} adopts a mixture of benchmarks to remove preference bias. NPHardEval~\cite{NPHardEval} takes inspiration from computational complexity theory to generate questions provably free from training data contamination through algorithmic construction rather than web scraping. Recent anti-leakage frameworks construct samples with explicitly new knowledge absent from LLMs' training sets~\cite{AntiLeakBench}. Nevertheless, contamination remains persistent when questions are sourced from the public Internet.

\paragraph{FTP for Contamination-Free Generation} Formal theorem proving offers a promising solution to contamination through mechanized verification using interactive theorem provers such as Lean, Isabelle, and Coq. Evaluation datasets like miniF2F~\cite{miniF2F} provide unified cross-system benchmarks with 488 Olympiad-level problems, while psychometric frameworks introduce difficulty grading reflecting theorem complexity as perceived by LLMs~\cite{PsychometricFTP}. LLMs have shown remarkable potential to advance mathematics through FTP~\cite{GenLMforATP}, endorsed by leading mathematicians such as Peter Scholze~\cite{LiquidTensor} and Terence Tao~\cite{TerryTao}. Large-scale synthetic approaches like DeepSeek-Prover demonstrate synthesis of millions of formal statements with proofs~\cite{DeepSeekProverV2}. Integration in LLM-based Agents~\cite{FormalAgent}, RAG systems~\cite{LeanRAG}, and applications like code copilots~\cite{CodeGeneration} and optimization~\cite{OptiBench} further demonstrate FTP's potential.

\paragraph{Automated Formal-Informal Pair Generation} Autoformalization automatically translates natural language into formal language, addressing the costly and challenging manual process. Recent advances employ theorem provers as judges to validate synthetic data generation through iterative approaches~\cite{TheoremProverJudge}, while methods generate arbitrary numbers of theorems across mathematical domains using Lean proof tactics. Previous synthetic data approaches~\cite{INT,MUSTARDSynthesis,LeanWorkbook} for LLM-based autoformalization~\cite{AutoFSymbolicEquiv,AutoFormalizationLLM,ProcessDrivenAutoF} remain limited by expert annotation requirements for proofreading LLM-generated formal content, which is labor-intensive and suffers from inter-annotator inconsistencies. This work aims to automate the entire (in)formalization process through templates ensuring rigorous alignment, rendering manual curation \textit{an optional quality check} rather than \textit{an expensive necessity}.
\section{Methodology}

Our approach employs a systematic three-stage synthesis framework that generates infinite families of formally rigorous mathematical challenges with provable correctness. The framework consists of: (1) \textbf{Problem Module} that defines parameterized computational problems with adjustable difficulty; (2) \textbf{Ground-Truth Generation} that employs algorithmic methods to determine correct answers; (3) \textbf{Template-Based Synthesis} that generates rigorously aligned formal-informal pairs using expert-defined templates. We demonstrate this pipeline through two exemplar instantiations: Busy Beaver machines testing undecidable halting problems, and Mixed Boolean-Arithmetic expressions testing symbolic algebraic manipulation, as illustrated in Figure~\ref{fig:SynthesisPipeline}.

\subsection{The Busy Beaver (BB) Challenge}\label{BBChallenge} 

The busy beaver game aims to find a terminating program of a given size that either produces the most output possible or runs for the longest number of steps, with busy beavers defined as Turing machines that perform a maximum number of steps when started on an initially blank tape before halting.  This problem, introduced by Rado in 1962, represents one of the most fundamental undecidable problems in computational theory. The profound mathematical significance extends far beyond theoretical computer science: recent breakthroughs have demonstrated that major mathematical conjectures, including the Collatz Conjecture and Goldbach Conjecture, reduce to specific Busy Beaver instances BB(6) and BB(27) respectively, establishing Busy Beaver problems as canonical representatives of mathematical undecidability.

\paragraph{Formal Definition of a BB Machine}
We define a Busy Beaver Turing machine as a 7-tuple $M = (Q, \Sigma, \Gamma, \delta, q_0, B, F)$ where: $Q = \{q_0, q_1, \ldots, q_{N-1}, q_{\text{HALT}}\}$ is a finite set of $N+1$ states; $\Sigma = \{0, 1\}$ is the binary input alphabet; $\Gamma = \Sigma$ is the tape alphabet; $\delta: (Q \setminus \{q_{\text{HALT}}\}) \times \Gamma \rightarrow Q \times \Gamma \times \{L, R\}$ is the transition function; $q_0$ is the initial state; $B = 0$ is the blank symbol; and $F = \{q_{\text{HALT}}\}$ is the set of final states. The machine operates on an infinite tape initially filled with zeros, where each step applies $\delta(q, s) = (q', s', d)$ to determine the next state $q'$, symbol $s'$ to write, and direction $d$ to move the head.

The Busy Beaver function is formally defined as $\textbf{BB}(N) = \max\{|\text{steps}(M)| : M \in \mathcal{H}_N\}$ where $\mathcal{H}_N$ denotes all $N$-state machines that halt on blank tape, and $|\text{steps}(M)|$ counts execution steps before halting. This function is undecidable because determining $\textbf{BB}(N)$ requires solving instances of the halting problem.  Each machine is specified by its state transition table with tuples $(\text{State}, \text{Symbol}, \text{Next State}, \text{Move}, \text{Write Symbol})$, as illustrated in Figure~\ref{fig:SynthesisPipeline}.

The parameter $N$ provides natural difficulty scaling through exponential growth in machine count, detailed in Table~\ref{tab:BB(N)}, ranging from 64 machines for $N=1$ to over 4 billion for $N=4$.

\begin{table}[h]
\centering
\begin{tabular}{cr}
\toprule
\( n \) & Number of Machines \\
\midrule
1 & 64 \\
2 & 20{,}736 \\
3 & 16{,}777{,}216 \\
4 & 4{,}294{,}967{,}296 \\
\bottomrule
\end{tabular}
\caption{Number of Busy Beaver (BB) Machines for Increasing Values of \( N \).}
\label{tab:BB(N)}
\end{table}

\paragraph{Challenge Task}
Models must prove whether a given $N$-state Busy Beaver machine $M$ (represented by its state transition table as in Figure~\ref{fig:SynthesisPipeline}) halts or runs indefinitely when initialized on a blank tape. This directly instantiates the classical halting problem in a concrete, mathematically well-defined setting requiring genuine logical reasoning.

\paragraph{Generating Easy-to-Verify Ground-Truth}
We leverage precisely known values of $\textbf{BB}(N)$ for small $N$ to establish provably correct classifications. Our algorithm executes each candidate machine for exactly $\textbf{BB}(N) + 1$ steps using deterministic simulation. Machines reaching $q_{\text{HALT}}$ within this bound are classified as halting; machines failing to halt are provably non-halting by the maximality property of the Busy Beaver function. This provides computationally efficient verification with mathematically rigorous classification boundaries.

\paragraph{Template-Based Synthesis for Hard-to-Prove Challenges}
Following Figure~\ref{fig:SynthesisPipeline}, we instantiate each machine through expert-designed templates generating both formal Lean theorem statements and natural language descriptions from identical state transition tables. The Lean template incorporates necessary definitions, complete machine specification, and theorem statement with \lstinline{sorry} placeholder for model proofs. The Markdown template provides structured natural language descriptions of the same mathematical objects, ensuring rigorous semantic alignment. The example in Figure~\ref{fig:SynthesisPipeline} demonstrates this dual representation maintaining mathematical precision across both modalities.

\begin{figure}
   \centering    \includegraphics[width=1\linewidth]{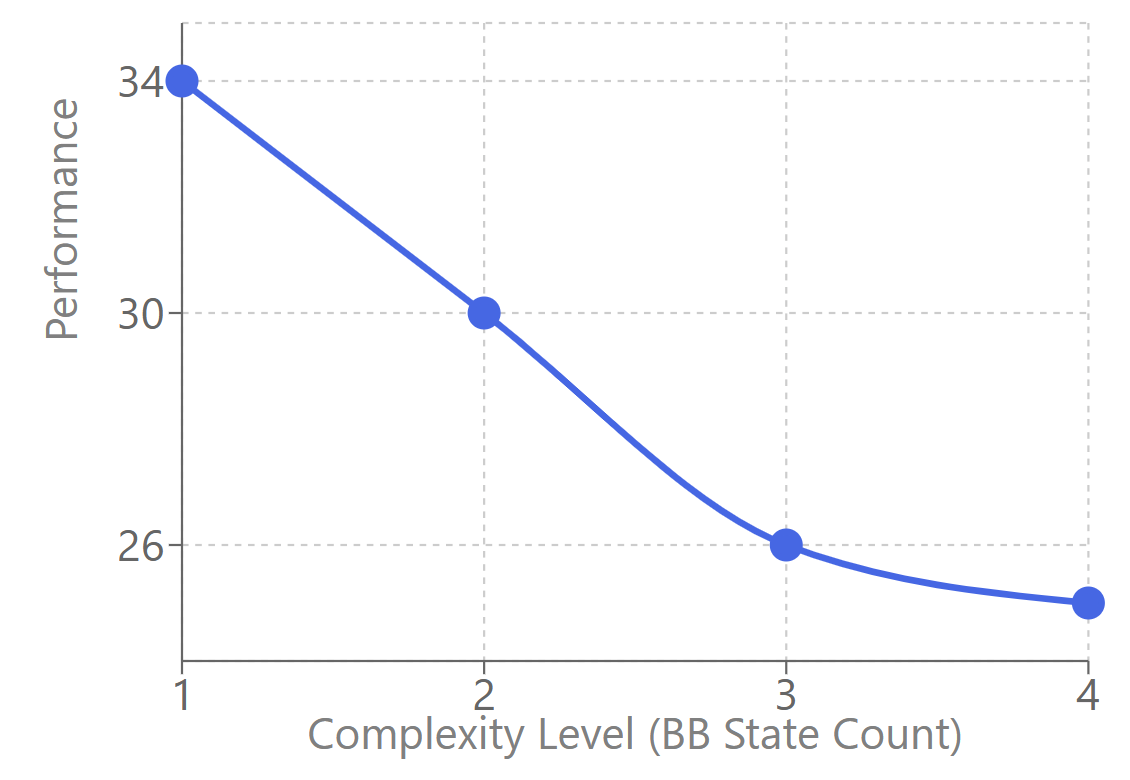}
   \caption{Performance degradation as BB complexity increases for best-performing DeepSeek-Prover-v2-671B}
\label{fig:performance2complexity}
\end{figure}

\subsection{The Mixed Boolean-Arithmetic (MBA) Challenge}\label{MBAChallenge}

Mixed Boolean Arithmetic (MBA) obfuscation transforms a program expression into an equivalent but complex expression that is hard to understand, performing semantics-preserving transformation from simple expressions to representations that are difficult to analyze.  MBA expressions strategically combine arithmetic operators $(+, -, \times)$ with bitwise operations $(\land, \lor, \oplus, \neg)$ over integer variables, creating expressions that use both conventional computer integer arithmetic on n-bit words and bitwise logic operations.  This technique finds extensive application in cryptographic systems, malware protection, and software obfuscation, where the goal is generating semantically equivalent but syntactically complex expressions that resist automated analysis while preserving computational correctness.

The mathematical foundation of MBA expressions operates on bitvectors $\mathbb{Z}_{2^w}$ where $w$ represents the word size (typically 32 or 64 bits). The fundamental insight exploits the fact that arithmetic and bitwise operations can be systematically combined to create equivalent expressions with vastly different syntactic representations. For example, simple addition $x + y$ can be obfuscated as $(x \oplus y) + 2 \times (x \land y)$, exploiting carry propagation properties of binary addition.

MBA equations manifest in two distinct structural categories:

\paragraph{Linear MBA Equations}
These equations take the form:
$$\underbrace{\sum_{k} a_k \times e_k(x,y)}_{\text{Linear MBA expression A}} = \underbrace{\sum_{k} b_k \times e_k'(x,y)}_{\text{Linear MBA expression B}}$$
where each coefficient $a_k, b_k \in \{\pm1, \pm2, \ldots, \pm11\}$ and each atomic boolean expression $e_k(x,y), e_k'(x,y)$ is drawn from the complete basis of 16 two-variable boolean combinations:
$$\left\{
\begin{array}{cccc}
x & x \lor y & x \land y & x \oplus y \\
\neg x & x \lor \neg y & x \land \neg y & \neg (x \oplus y) \\
y & \neg (x \lor y) & \neg x \land y & \neg (x \land y) \\
\neg y & \neg (x \lor \neg y) & \neg (x \land \neg y) & \neg (\neg x \land y)
\end{array}
\right\}$$

\paragraph{Nonlinear MBA Equations}
These involve products of MBA sub-expressions $\bigl(e_{k_1}(x,y) \times e_{k_2}(x,y)\bigr)$ and arbitrarily nested bitwise operations, such as:
$$10 \times \neg\!\Bigl(1\times(x\land y)\;\lor\;\neg\bigl((-1)\times(x\lor\neg y)\bigr)\Bigr)$$
This hierarchical structure creates multiple levels of algebraic complexity requiring sophisticated symbolic manipulation techniques.

\paragraph{Dataset Generation Following Synthesis Pipeline}
Following the established MBA obfuscator protocol and our synthesis pipeline in Figure~\ref{fig:SynthesisPipeline}, we construct equations by first generating linear MBA expressions that are identically zero, then randomly redistributing terms across equation sides with appropriately flipped signs to ensure semantic equivalence. Each equation translates into a Lean theorem paired with an informal Markdown description as demonstrated in Figure~\ref{fig:SynthesisPipeline}. All variables and constant coefficients are declared as \texttt{BitVec 32}, ensuring arithmetic and bitwise operations operate on 32-bit bitvectors modulo $2^{32}$. The parameter space scales infinitely by incrementing variable numbers in MBA expressions, guaranteeing sufficient problem diversity through combinatorial explosion.

The specific MBA equation shown in Figure~\ref{fig:SynthesisPipeline} exemplifies this approach: variables $x, y$ are 32-bit bitvectors with the theorem requiring step-by-step proof that two syntactically complex but semantically equivalent MBA expressions represent identical computational functions. The formal Lean statement precisely captures bitwise arithmetic semantics while the informal description explains the same mathematical relationship in natural language.

\paragraph{Challenge Task}
Models must prove equivalence between two MBA expressions through explicit step-by-step symbolic reasoning using fundamental algebraic identities, bitwise operation laws, and distributive properties. We explicitly prohibit automatic tactics such as \texttt{bv\_decide} (which invokes external SAT solvers) to ensure genuine symbolic manipulation rather than computational verification, thereby testing authentic theorem proving capabilities rather than automated decision procedures.

\paragraph{Generating Easy-to-Verify Ground-Truth via Weighted 2-DNF Canonical Forms}
To establish verifiable ground-truth for MBA equation equivalence, we employ a systematic normalization process that converts any MBA expression into a unique canonical representation. For any linear MBA equation $E_1(x,y) = E_2(x,y)$ over two variables, we normalize both sides into their Weighted 2-Disjunctive Normal Form (W2DNF). This canonical form represents any boolean-arithmetic expression as a weighted sum of all possible conjunctions of the two variables and their negations:
$$\mathrm{W2DNF}(E) = \sum_{i,j\in\{0,1\}} c_{ij}(E) \times (\ell_i(x) \land \ell_j(y))$$
where the literal functions are defined as $\ell_0(z) = \neg z$ and $\ell_1(z) = z$, creating four fundamental conjunctive terms: $\neg x \land \neg y$, $\neg x \land y$, $x \land \neg y$, and $x \land y$. Each coefficient $c_{ij}(E) \in \mathbb{Z}_{2^{32}}$ represents the weight of the corresponding conjunctive term and is uniquely determined by the structure of expression $E$.

The key mathematical property is that two MBA expressions $E_1$ and $E_2$ are semantically equivalent if and only if their W2DNF representations have identical coefficients: $c_{ij}(E_1) = c_{ij}(E_2)$ for all $i,j \in \{0,1\}$. This provides an algorithmic decision procedure: we compute the four coefficients for each expression and verify their pairwise equality. For the concrete example shown in Figure~\ref{fig:SynthesisPipeline}, both complex MBA expressions on the left and right sides of the equation reduce to identical W2DNF coefficient vectors, thereby confirming their mathematical equivalence through systematic algebraic normalization.

We implement this normalization process by systematically applying algebraic identities to reduce each MBA expression to its canonical W2DNF form, then translating these reduction steps directly into sequences of verifiable Lean tactics using our curated lemma library. This approach provides complete ground-truth solutions that models must replicate through symbolic reasoning, ensuring that successful proof construction demonstrates genuine understanding of bitwise-arithmetic identities rather than computational guessing.

\paragraph{Customized Lemma Library}
We developed a comprehensive custom library containing all necessary algebraic identities to reduce any 2-variable linear MBA expression into W2DNF canonical form without external theorem dependencies. This library includes fundamental laws such as distributivity of arithmetic over bitwise operations, De Morgan's laws for bitwise negation, and specialized identities like $x \oplus y = x + y - 2 \times (x \land y)$ that connect bitwise and arithmetic operations. The library enables automatic generation and verification of all ground-truth proofs while ensuring model solutions depend solely on fundamental mathematical reasoning rather than domain-specific automation.

\paragraph{Step-Level Task Decomposition}
To evaluate fine-grained reasoning capabilities beyond complete proof construction, we decompose each ground-truth proof into sequences of individual step-level tasks corresponding to the systematic transformation process illustrated in Figure~\ref{fig:SynthesisPipeline}. After each tactic application in our ground-truth proofs, Lean reports updated proof goals, and whenever the left-hand side (LHS) or right-hand side (RHS) of the current goal changes, we extract a corresponding step lemma following systematic rules. For LHS changes, the extracted lemma has its LHS matching the previous goal's LHS and its RHS matching the current goal's LHS; for RHS changes, the lemma's LHS matches the previous goal's RHS and its RHS matches the current goal's RHS. Each problem's final step employs the \texttt{simp} tactic for goal completion, while all non-final tactics exclusively use \texttt{simp only}, \texttt{rw}, or \texttt{nth\_rewrite} with lemmas from our custom library. This decomposition ensures that every individual reasoning step can be independently verified and evaluated, enabling detailed analysis of model capabilities at the granular level of single algebraic transformations.
\section{Experiments}\label{Exp}
We report Pass@$16$ on various frontier models, which counts a case as successful if and only if at least one of the $16$ attempts could pass Lean compiler verification, that is, no errors are reported and no 'sorry' is used in the proof. 

While it's possible to set other $n$ for Pass@$n$, we note that $n=16$ is the common best practice for existing formal theorem proving benchmarks~\cite{miniF2F,PutnamBench,CombiBench}, which strikes a balance of reasonable budget and non-deterministic inference. 
For unexpected failures such as Timeout, Out-Of-Memory and API error, we re-run each trial without counting the failed case into $n=16$.

\begin{table*}[ht]
    \centering
    \begin{tabular}{
        lccccc
    }
        \toprule
         \textbf{Model} 
            & \textbf{BB(1)} 
            & \textbf{BB(2)}
            & \textbf{BB(3)}
            & \textbf{BB(4)}
            & \textbf{Total} \\
        \midrule
        \textit{General LRMs} &&&&& \\ 
        \midrule
        DeepSeek-R1 
            & 21/50 & 12/50 & 10/50 & 12/50 & 55/200 (27.5\%) \\
        QwQ-32B-Preview 
            & 26/50 & 21/50 & 26/50 & 25/50 & 98/200 (49.0\%) \\
        OpenAI-o3 
            & 15/50 & 17/50 & 19/50 & 17/50 & 68/200 (34.0\%) \\
        OpenAI-o4-mini 
            & 19/50 & 10/50 & 12/50 & 13/50 & 54/200 (27.0\%) \\
        \midrule
        \textit{Theorem Provers} &&&&& \\
        \midrule        
        Goedel-Prover-SFT 
            & 0/50 & 0/50 & 0/50 & 0/50 & 0/200 (0.0\%) \\
        Leanabell-Prover-GD-RL 
            & 0/50 & 0/50 & 0/50 & 0/50 & 0/200 (0.0\%) \\
        Kimina-Prover-Preview-7B 
            & 0/50 & 0/50 & 0/50 & 0/50 & 0/200 (0.0\%) \\
        DeepSeek-Prover-v2-7B 
            & 9/50 & 9/50 & 10/50 & 10/50 & 38/200 (19.0\%) \\
        DeepSeek-Prover-v2-671B 
            & 34/50 & 30/50 & 26/50 & 25/50 & 115/200 (57.5\%) \\
        \midrule
    \end{tabular}
    \caption{
        Pass@16 Results on BB Challenge, where each column represents an incremental setting for state count $N$ as a complexity parameter.
    }
    \label{tab:res-whole-proof-bb}
\end{table*}

\subsection{Evaluation Setup}

\paragraph{Evaluated models.}
We evaluated both frontier general-purpose reasoning models and LLM-based theorem provers. 
Dedicated formal theorem provers are fine-tuned using FTP data on base models such as DeepSeekMath~\cite{DeepSeekMath}.
For general-purpose LRMs, we chose a mixture of advanced open-source and proprietary models: DeepSeek-R1-671B \cite{DeepSeekR1}, QwQ-32B~\cite{QwQ32B}, OpenAI-o3 and o4-mini~\cite{o3}. Both OpenAI models carry default settings of 200K context window, 100K max output tokens, and a knowledge cut-off date on June 1, 2024.
For LLM-based theorem provers, we evaluated various frontier models including Goedel-Prover-SFT~\cite{Goedelprover}, Leanabelle-Prover~\cite{Leanabelle}, Kimina-Prover-Preview-7B~\cite{KiminaProver}, DeepSeek-Prover-v2-7B, and DeepSeek-Prover-v2-671B~\cite{DeepSeekProverV2}.

\paragraph{Inference setting.}
For open-source models, we use greedy sampling to aid reproducibility, which is equivalent to fixing the sampling temperature to $T=0.0$ as recommended by the model developers~\cite{DeepSeekR1,QwQ32B} for math and coding tasks. 
Furthermore, We enable thinking mode for all models whenever possible and follow default settings on Reasoning Efforts, Context Length, Max Tokens, and System Prompts in model metadata.

We implemented our code under a Linux environment with \texttt{Python 3.10}. 
All open-source models are deployed using \texttt{transformer} library and \texttt{PyTorch} on a 4xGH200 server, with open-source models downloaded via \texttt{huggingface-cli}. 
The proprietary models are used directly via the OpenAI API.
We prompted models to follow detailed instruction as in Appendix~\ref{prompt}.

After generation, model outputs are parsed to extract Lean proof, which is validated using a regular expression. 
The proof is then verified using \texttt{Kimina-Lean-Server}~\cite{KiminaLeanServer} built with \texttt{Lean and Mathlib v4.19.0} via a RESTful API.

\subsection{Results}
\paragraph{BB challenge.}
We report Pass@$16$ in Table~\ref{tab:res-whole-proof-bb} and analyze detailed failure modes in Section~\ref{FailureCase}. 
Frontier LRMs OpenAI-o3 and o4-mini have struggled to score past~35\% on our BB challenge, marking a critical gap in formal reasoning capability despite rapid saturation on static benchmarks from public sources.

Among LLM-based theorem provers, best-performing DeepSeek-Prover-v2-671B~\cite{DeepSeekProverV2} scored~57.5\%, but we notice rapid performance degradation as models scale down to 7B, where advanced theorem provers such as Goedel-Prover-SFT \cite{Goedelprover}, Leanabelle-Prover-GD-RL~\cite{Leanabelle} and Kimina-Prover-Preview-7B~\cite{KiminaProver} struggled to solve even 1 instance of our challenges.

Paradoxically, QwQ-32B~\cite{QwQ32B} attained best performance among general LRMs with only 32B parameters, which raises an intriguing question as to how scaling affects the performance of general reasoning models vs. dedicated theorem provers in a potentially different manner.

\paragraph{MBA challenge.}
We report Pass@$16$ in Table~\ref{tab:res-mba-proof}, revealing even more pronounced difficulties than BB challenge. Further, we decompose 10 randomly selected MBA challenge into 365 steps and report step-level performance in Table~\ref{tab:step-res-mba-proof}.

Frontier LRMs including DeepSeek-R1~\cite{DeepSeekR1} and QwQ-32B~\cite{QwQ32B}, struggled to solve any of our challenges, while OpenAI-o4-mini~\cite{o3} solved only 6 problems.

Specialized theorem provers exhibited extreme stratification: DeepSeek-Prover-v2-671B~\cite{DeepSeekProverV2} solved~12\% of linear problems with lemmas, but failed all problems without lemmas and all non-linear problems. Smaller provers showed near-universal failure, with only Kimina-Prover-7B~\cite{KiminaProver} solving a single instance. This demonstrates that even state-of-the-art formal systems struggle with fundamental algebraic reasoning when deprived of lemma support or faced with non-linear complexity.

\begin{table*}[ht]
    \centering
    \begin{tabular}{lccc}
        \toprule
        \textbf{Model} & \textbf{Linear w/ Lemma} & \textbf{Linear w/o Lemma} & \textbf{Non-Linear} \\
        \midrule
        \textit{Reasoning LLMs} \\
        \midrule
        DeepSeek-R1 
        & 0/100 & 0/100 & 0/40 \\
        OpenAI-o4-mini 
        & 2/100 & 4/100 & 0/40 \\
        QwQ-32B-Preview 
        & 0/100 & 0/100 & 0/40 \\
        \midrule
        \textit{Theorem Provers} \\
        \midrule
        DeepSeek-Prover-v2-7B 
        & 0/100 & 0/100 & 0/40 \\
        Goedel-Prover 
        & 0/100 & 0/100 & 0/40 \\
        Kimina-Prover-7B 
        & 1/100 & 0/100 & 0/40 \\
        Leanabelle-Prover 
        & 0/100 & 0/100 & 0/40 \\
        DeepSeek-Prover-v2-671B 
        & 12/100 & 0/100 & 0/40 \\
        \bottomrule
    \end{tabular}
    \caption{
        Pass@$16$ Performance on MBA challenge. We evaluated all models using 2 parallel settings where models are tested with vs. without our customized lemma library as detailed in Appendix~\ref{custom lemma library}. This parallel setting aims to gauge their capability of in-context learning unseen lemma to tackle novel problems out of their training corpora.
    }
    \label{tab:res-mba-proof}
\end{table*}

\begin{table}[ht]
    \centering
    \begin{tabular}{lcc}
        \toprule
        \textbf{Model} & \textbf{\#Correct Steps} \\
        \midrule
        OpenAI-o4-mini & 352/356 \\
        DeepSeek-Prover-v2-671B & 258/356 \\
        \bottomrule
    \end{tabular}
    \caption{
         Step-level task performance on linear MBA where models are asked to choose an appropriate lemma from our customized lemma library for each step of the proof.
    }
    \label{tab:step-res-mba-proof}
\end{table}

\subsection{Success Cases}
We present a successful case of a non-halting BB(4) machine and show more case study in Appendix~\ref{AdditionalCase}.
The model DeepSeekProver-v2-671B leveraged proof by induction and attempted various automated tactics, such as \texttt{aesop} and \texttt{omega}, showing that a considerable manipulation of the automated tactic taxonomy.

\begin{lstlisting}[breaklines]
DeepSeek-Prover-v2-671B:
theorem machine_never_halts : ∀ n, (nth_cfg n).isSome := by
  intro n
  have h : ∀ n, (nth_cfg n).isSome := by
    intro n
    induction n with
    | zero => simp [nth_cfg, init]
    | succ n ih =>
      cases h₁ : nth_cfg n with
      | none => simp_all [nth_cfg, step, machine, Cfg, Stmt, Γ, Fin.forall_fin_succ, Fin.forall_fin_succ']
      | some cfg =>
        simp_all [nth_cfg, step, machine, Cfg, Stmt, Γ, Fin.forall_fin_succ, Fin.forall_fin_succ']
        <;>
        (try cases cfg <;> simp_all [nth_cfg, step, machine, Cfg, Stmt, Γ, Fin.forall_fin_succ, Fin.forall_fin_succ'])
        <;>
        (try aesop)
        <;>
        (try omega)
        <;>
        (try aesop)
  exact h n
\end{lstlisting}

In this successful MBA case, the problem is simple enough that \texttt{bv\_decide} is able to solve it within its default 10-second timeout by leveraging a SAT solver.
\begin{lstlisting}[breaklines]
o4-mini:
theorem mba_challenge_88282d89 (x y : BitVec 32) : 1#32 * ~~~(x &&& ~~~y) + 1#32 * ~~~(x ^^^ y) - 3#32 * (x ||| ~~~y) + 1#32 * ~~~(x ||| y) + 3#32 * (x &&& ~~~y) + 1#32 * (x &&& y) = 1#32 * ~~~(x ||| ~~~y) := by
  bv_decide
\end{lstlisting}

This case is more complex such that \texttt{bv\_decide} alone cannot solve. However, the combination of \texttt{norm\_num} and \texttt{bv\_decide} is sufficient to complete the proof. Notably, the sequence of have statements introducing previously proven lemmas is entirely unused, and the tactics following the first \texttt{bv\_decide} are never executed, as the goal is already solved at that point. This suggests that the model remains heavily reliant on automated tactics and lacks a clear understanding of what these tactics actually do.

\begin{lstlisting}[breaklines]
DeepSeek-Prover-v2-671B:
theorem mba_challenge_3165c251 (x y : BitVec 32) :  
  -3#32 * (x ||| ~~~y) - 1#32 * x - 2#32 * y + 1#32 * ~~~(x ||| y) + 2#32 * ~~~(x ||| ~~~y) + 2#32 * (x &&& ~~~y) + 6#32 * (x &&& y) = -2#32 * ~~~y := by
  have h₁ := bv32_not_xor_eq_or x y
  ...
  have h₂₁ := bv32_sum_all y x
  norm_num [Nat.add_assoc, Nat.add_comm, Nat.add_left_comm, Nat.mul_assoc, Nat.mul_comm, Nat.mul_left_comm] at *
  <;> simp_all (config := {decide := true})
  <;> bv_decide
  <;> simp_all (config := {decide := true})
  <;> bv_decide
\end{lstlisting}

\subsection{Failure Cases}\label{FailureCase}
Our experiments reveal several key failure modes of frontier models.
These failure can be categorized into 4 major types with signature patterns as shown in Table~\ref{tab:res-mba-proof}. More cases in Appendix~\ref{AdditionalCase}.

\begin{table}[htb]
    \centering
    \begin{tabular}{ll}
        \toprule
        \textbf{Error Message Type} & \textbf{\%}(\#Errors) \\
        \midrule
Irrelevant Hallucination & 67.27\% (122404) \\
Tactic Misuse & 23.22\% (42247) \\
Voluntary Give Up & 4.88\% (8871) \\
Type Mismatch & 4.47\% (8136) \\
Others & 0.16\% (300) \\
        \bottomrule
    \end{tabular}
    \caption{Summary of error type distribution for both BB and MBA Challenge.}
    \label{tab:ErrorType}
\end{table}

\paragraph{Hallucination.}
Hallucination consistently appears within 7B theorem provers, where they often tried to use non-existent theorems or tactics. DeepSeek-Prover-v2-671B has exhibited much stronger anti-hallucination capability thanks to its large base model. 

Here we show 2 cases where Goedel-Prover-SFT attempted to prove by induction on a completely unrelated problem. Similarly, Kimina-Prover-Preview tried to rename the given theorem head \texttt{nth\_cfg} to \texttt{nth\_configuration}, which led to an unsolved goal at the end of the proof.

\begin{lstlisting}[breaklines]
Goedel-Prover-SFT:
theorem sequence_formula {a : ℕ → ℕ} (h1 : a 1 = 1)
    (h_rec : ∀ n : ℕ, n ≥ 1 → a (n + 1) = 2 * a n + 1) :
    ∀ n : ℕ, n ≥ 1 → a n = 2^n - 1 := by
\end{lstlisting}

\begin{lstlisting}[breaklines]
Kimina-Prover-Preview:
theorem machine_never_halts : ∀ n, (nth cfigurations n).isSome := by
  intro n
  induction n with
  | zero =>
    simp [nth configurations]
  | succ n ih =>
    simp [nth configurations, ih]
    <;> aesop
\end{lstlisting}

For the MBA challenge, we show another failure case of OpenAI-o4-mini where model hallucinates a non-existent tactic \texttt{bv\_simp}.
\begin{lstlisting}[breaklines]
o4-mini:
theorem mba_challenge_14b066fa (x y : BitVec 32) : -1#32 * ~~~(x ^^^ y) + 2#32 * ~~~x + 2#32 * ~~~(x &&& y) + 1#32 * (x ||| y) - 3#32 * x + 3#32 * (x ||| ~~~y) - 2#32 * (x ^^^ y) + 1#32 * ~~~(x &&& ~~~x) - 5#32 * y - 8#32 * ~~~(x ||| y) - 2#32 * (x &&& ~~~y) + 3#32 * (x &&& y) = -1#32 * ~~~(x &&& ~~~y) := by 
    bv_simp 
    ring
\end{lstlisting}

\paragraph{Tactic misuse.}
Models showed a dangerous tendency to blindly apply automated proof tactics such as aesop (which automatically looks for search tactics recursively in a proof tree) without understanding the necessary conditions for aesop to be applicable, resulting in the failure of exhaustive search.

\begin{lstlisting}[breaklines]
DeepSeek-Prover-v2-671B:
theorem machine_never_halts : ∀ n, (nth configurations n).isSome := by
  intro n
  ...
    <;> aesop
\end{lstlisting}

This tendency is even more pronounced in the following examples. The model gathers many lemmas into one call without checking their relevance and then blindly invokes ring. The proof is bloated with invalid rewrites and heavy automation, yet still fails，demonstrating a systematic failure to reason about lemma and tactic applicability.
\begin{lstlisting}[breaklines]
o4-mini with lemmas:
theorem mba_challenge_080db093 (x y : BitVec 32) :
  -5#32 * ~~~y + 1#32 * ~~~(x &&& ~~~x) + 4#32 * (x ||| y) - 6#32 * (x ^^^ y) - 6#32 * ~~~(x &&& ~~~y) - 1#32 * ~~~(x &&& y) - 3#32 * y - 11#32 * ~~~x + 24#32 * ~~~(x ||| y) + 22#32 * ~~~(x ||| ~~~y) + 13#32 * (x &&& ~~~y) +  6#32 * (x &&& y) = -4#32 * ~~~(x ^^^ y) + 6#32 * (x ||| ~~~y) := by
  simp only [bv32_and_not_self, bv32_not_and_self, bv32_or_not_self, bv32_not_or_self] at *
  simp only [bv32_x_distr, bv32_y_distr, ...] at *
  simp only [bv32_sum_all] at *
  ring

\end{lstlisting}

This sample illustrates a tactic misuse where\texttt{bv\textunderscore decide} fails to solve the goal within the 10-second timeout. The failure highlights the limitations of relying solely on automated tactics for large or intricate goals.
\begin{lstlisting}
o4-mini:
import Std.Tactic.BVDecide
/-- 
Let x,y be 32-bit bit-vectors. Prove the equivalence of the following two expressions: 7·x - 5·y - 2·(x⊕y) - 6·¬(x ∧ ¬x) - 5·(x∨y) - 2·¬(x∧y) - (x∨¬y) + 4·¬y - 7·¬(x⊕y) + 13·¬(x∨y) + 21·¬(x∨¬y) + 17·(x∧y) and -5·(x∧¬y) + 1·¬x 
-/
theorem mba_challenge_6f99807f (x y : BitVec 32) : ((7#32 * x) - (5#32 * y) - (2#32 * (x ^^^ y)) - (6#32 * (~~~(x &&& (~~~x)))) - (5#32 * (x ||| y)) - (2#32 * (~~~(x &&& y))) - (1#32 * (x ||| (~~~y))) + (4#32 * (~~~y)) - (7#32 * (~~~(x ^^^ y))) + (13#32 * (~~~(x ||| y))) + (21#32 * (~~~(x ||| (~~~y)))) + (17#32 * (x &&& y))) = ((-5#32) * (x &&& (~~~y))) + (1#32 * (~~~x)) := by
  bv_decide
\end{lstlisting}

\paragraph{Type mismatch.}
Type mismatch is a fundamental syntactic error arising from the type-dependent nature of Lean system, which directly reflects models' failure to understand and correctly apply respective Lean tactics in question. In this case, the 7B theorem-proving models failed to interpret and adapt to Lean’s inherent type rules, producing expressions whose inferred types did not match the expected ones. Fortunately, this type of error occurs rarely at 4.47\% in the big picture, which speaks to the promising progress that frontier models have a considerably correct understanding of the Lean tactic taxonomy.

\begin{lstlisting}[breaklines]
Kimina-Prover-Preview:
theorem machine_never_halts : ∀ n, (nth_cfg n).isSome := by
  intro n
  induction n with
    ...
  | succ n ih =>
    simp [nth_cfg, step, machine]
    exact ih
\end{lstlisting}

\paragraph{Voluntary give-up.}
In this case, the model voluntarily chose to give up by leaving \texttt{sorry} under the proof goal without even attempting to solve the problem with any tactics at all. This type of error makes up 4.88\% of total errors, which could be seen as a failure of instruction-following often depicted by reasoning models comparing to general-purpose LLMs.

\begin{lstlisting}[breaklines]
Kimina-Prover-Preview:
theorem machine_never_halts : ∀ n, (nth_cfg n).isSome := by
  sorry
\end{lstlisting}


\section{Discussion}

\paragraph{The Automation Trap in Theorem Proving}
Our empirical analysis identifies a systematic overreliance on automated tactics as primary proof mechanisms across frontier models. Models predominantly deploy tactics such as \texttt{aesop} for recursive proof search and \texttt{bv\_decide} for SAT-based solving as black-box oracles, demonstrating minimal comprehension of their formal operational boundaries. This dependency manifests in two critical failure modalities: (1) proofs that circumvent genuine deductive understanding when tactics succeed coincidentally, and (2) complete proof collapse when tactics fail without fallback mechanisms. Notably, models frequently generate elaborate lemma declarations while defaulting to brute-force tactic sequencing—a pattern indicating a fundamental disconnect between syntactic manipulation and strategic reasoning that fundamentally undermines verification integrity.

\paragraph{The Reasoning Gap between Step-Level vs. Holistic Proof Task}
Our step-level task decomposition setting reveals a significant discrepancy between step-level performance and holistic proof synthesis with a simple overhead proof goal. Models demonstrate near-perfect capability in choosing unseen out-of-distribution lemma for  atomic inference tasks yet exhibit catastrophic failure rates in composing these operations into complete proofs. This divergence indicates that the primary bottleneck resides not in local operations but in global proof planning and strategic integration. The persistent inability to reconcile stepwise correctness with end-to-end whole proof generation suggests reasoning capabilities of frontier models remain constrained in long contexts by insufficient strategic planning for orchestrating atomic transformations into coherent proof strategies when challenged with lengthy proofs containing multiple steps.

\paragraph{Contamination-Resistant Evaluation Framework}
Our methodology establishes a novel evaluation paradigm through the integration of Lean with Theoretical Computer Science. Our framework achieves infinite scalability via algorithmic problem generation from parameterized TCS modules, with granular difficulty modulation through computational parameters (e.g., Turing Machine state complexity, MBA expression depth). Automated verifiability provides ground-truth validation without human intervention, while dynamically generated problem spaces ensure intrinsic resistance to dataset contamination. By enabling rigorous formal-informal alignment absent expert curation, this TCS-inspired synthesis creates a sustainable evaluation ecosystem where benchmark freshness and complexity scales along with the progress of frontier theoretical computer science research.

\section{Conclusion}
We propose Theoretical Computer Science as a promising testbed for formal theorem proving, which by design is capable of utilizing modularized, fully automatic problem generation to systematically overcome contamination and scalability limitations in legacy benchmarks. Our TCS-inspired modular synthesis paradigm enables the creation of infinitely scalable, formally rigorous problem spaces with adjustable complexity through computational parameters such as Turing Machine state configurations and MBA expression depth. Our experimental results demonstrate a profound reasoning gap: while models achieve near-perfect scores of 98.88\% on atomic step-level tasks involving out-of-distribution lemma selection, they collapse to a mere 12\% success rate when synthesizing complete lengthy proofs. This drastic performance degradation highlights the devastating role of hallucination during long-context reasoning sequences, which may act as a roadblock for models to form any systematic strategies for tasks requiring lengthy whole proofs. We also reveal the dangerous tendency of frontier models' over-reliance on automated tactics like \texttt{aesop} and \texttt{bv\_decide} without understanding their applicability with limitations. To tackle these challenges, we call on the Lean community to further explore the massive potential of theoretical computer science and its interplay with the realm of formal theorem proving, thereby leveraging the joint effort to advance both domains synergistically.

\newpage
\bibliography{References}

@misc{DeepSeekMath,
      title={DeepSeekMath: Pushing the Limits of Mathematical Reasoning in Open Language Models}, 
      author={Zhihong Shao and Peiyi Wang and Qihao Zhu and Runxin Xu and Junxiao Song and Xiao Bi and Haowei Zhang and Mingchuan Zhang and Y. K. Li and Y. Wu and Daya Guo},
      year={2024},
      eprint={2402.03300},
      archivePrefix={arXiv},
      primaryClass={cs.CL},
      url={https://arxiv.org/abs/2402.03300}, 
}

@misc{Leanabelle,
      title={Leanabell-Prover: Posttraining Scaling in Formal Reasoning}, 
      author={Jingyuan Zhang and Qi Wang and Xingguang Ji and Yahui Liu and Yang Yue and Fuzheng Zhang and Di Zhang and Guorui Zhou and Kun Gai},
      year={2025},
      eprint={2504.06122},
      archivePrefix={arXiv},
      primaryClass={cs.AI},
      url={https://arxiv.org/abs/2504.06122}, 
}

@misc{QwQ32B,
  title = {QwQ - 32B: The Power of Scaling RL},
  url = {https://qwenlm.github.io/blog/qwq - 32b/},
  author = {Qwen},
  month = {March},
  year = {2025},
  note = {Accessed: 2025 - 05 - 14}
}

@article{FTPReview,
  author       = {Si Chen and
                  Wensheng Yu and
                  Guowei Dou and
                  Qimeng Zhang},
  title        = {A Review on Mechanical Proving and Formalization of Mathematical Theorems},
  journal      = {{IEEE} Access},
  volume       = {13},
  pages        = {50672--50686},
  year         = {2025},
  url          = {https://doi.org/10.1109/ACCESS.2025.3552634},
  doi          = {10.1109/ACCESS.2025.3552634},
  bibsource    = {dblp computer science bibliography, https://dblp.org}
}

@inproceedings{AutoFSymbolicEquiv,
  author       = {Zenan Li and
                  Yifan Wu and
                  Zhaoyu Li and
                  Xinming Wei and
                  Xian Zhang and
                  Fan Yang and
                  Xiaoxing Ma},
  editor       = {Amir Globersons and
                  Lester Mackey and
                  Danielle Belgrave and
                  Angela Fan and
                  Ulrich Paquet and
                  Jakub M. Tomczak and
                  Cheng Zhang},
  title        = {Autoformalize Mathematical Statements by Symbolic Equivalence and
                  Semantic Consistency},
  booktitle    = {Advances in Neural Information Processing Systems 38: Annual Conference
                  on Neural Information Processing Systems 2024, NeurIPS 2024, Vancouver,
                  BC, Canada, December 10 - 15, 2024},
  year         = {2024},
  url          = {http://papers.nips.cc/paper\_files/paper/2024/hash/6034a661584af6c28fd97a6f23e56c0a-Abstract-Conference.html},
  bibsource    = {dblp computer science bibliography, https://dblp.org}
}

@article{ProcessDrivenAutoF,
  author       = {Jianqiao Lu and
                  Zhengying Liu and
                  Yingjia Wan and
                  Yinya Huang and
                  Haiming Wang and
                  Zhicheng Yang and
                  Jing Tang and
                  Zhijiang Guo},
  title        = {Process-Driven Autoformalization in Lean 4},
  journal      = {CoRR},
  volume       = {abs/2406.01940},
  year         = {2024},
  url          = {https://doi.org/10.48550/arXiv.2406.01940},
  doi          = {10.48550/ARXIV.2406.01940},
  eprinttype    = {arXiv},
  eprint       = {2406.01940},
  bibsource    = {dblp computer science bibliography, https://dblp.org}
}

@misc{OptiBench,
      title={OptiBench Meets ReSocratic: Measure and Improve LLMs for Optimization Modeling}, 
      author={Zhicheng Yang and Yiwei Wang and Yinya Huang and Zhijiang Guo and Wei Shi and Xiongwei Han and Liang Feng and Linqi Song and Xiaodan Liang and Jing Tang},
      year={2024},
      eprint={2407.09887},
      archivePrefix={arXiv},
      primaryClass={cs.LG},
      url={https://arxiv.org/abs/2407.09887}, 
}

@inproceedings{Contam1,
  author       = {Inbal Magar and
                  Roy Schwartz},
  editor       = {Smaranda Muresan and
                  Preslav Nakov and
                  Aline Villavicencio},
  title        = {Data Contamination: From Memorization to Exploitation},
  booktitle    = {Proceedings of the 60th Annual Meeting of the Association for Computational
                  Linguistics (Volume 2: Short Papers), {ACL} 2022, Dublin, Ireland,
                  May 22-27, 2022},
  pages        = {157--165},
  publisher    = {Association for Computational Linguistics},
  year         = {2022},
  url          = {https://doi.org/10.18653/v1/2022.acl-short.18},
  doi          = {10.18653/V1/2022.ACL-SHORT.18},
  bibsource    = {dblp computer science bibliography, https://dblp.org}
}

@inproceedings{Contam2,
  author       = {Yihong Dong and
                  Xue Jiang and
                  Huanyu Liu and
                  Zhi Jin and
                  Bin Gu and
                  Mengfei Yang and
                  Ge Li},
  editor       = {Lun{-}Wei Ku and
                  Andre Martins and
                  Vivek Srikumar},
  title        = {Generalization or Memorization: Data Contamination and Trustworthy
                  Evaluation for Large Language Models},
  booktitle    = {Findings of the Association for Computational Linguistics, {ACL} 2024,
                  Bangkok, Thailand and virtual meeting, August 11-16, 2024},
  pages        = {12039--12050},
  publisher    = {Association for Computational Linguistics},
  year         = {2024},
  url          = {https://doi.org/10.18653/v1/2024.findings-acl.716},
  doi          = {10.18653/V1/2024.FINDINGS-ACL.716},
  bibsource    = {dblp computer science bibliography, https://dblp.org}
}

@inproceedings{ContaminationDetection,
  author       = {Jasper Dekoninck and
                  Mark Niklas M{\"{u}}ller and
                  Martin T. Vechev},
  editor       = {Amir Globersons and
                  Lester Mackey and
                  Danielle Belgrave and
                  Angela Fan and
                  Ulrich Paquet and
                  Jakub M. Tomczak and
                  Cheng Zhang},
  title        = {ConStat: Performance-Based Contamination Detection in Large Language
                  Models},
  booktitle    = {Advances in Neural Information Processing Systems 38: Annual Conference
                  on Neural Information Processing Systems 2024, NeurIPS 2024, Vancouver,
                  BC, Canada, December 10 - 15, 2024},
  year         = {2024},
  url          = {http://papers.nips.cc/paper\_files/paper/2024/hash/a7f89793b9e6f8c6568dbbb6ff727b9b-Abstract-Conference.html},
  bibsource    = {dblp computer science bibliography, https://dblp.org}
}

@misc{KiminaProver,
      title={Kimina-Prover Preview: Towards Large Formal Reasoning Models with Reinforcement Learning}, 
      author={Haiming Wang and Mert Unsal and Xiaohan Lin and Mantas Baksys and Junqi Liu and Marco Dos Santos and Flood Sung and Marina Vinyes and Zhenzhe Ying and Zekai Zhu and Jianqiao Lu and Hugues de Saxcé and Bolton Bailey and Chendong Song and Chenjun Xiao and Dehao Zhang and Ebony Zhang and Frederick Pu and Han Zhu and Jiawei Liu and Jonas Bayer and Julien Michel and Longhui Yu and Léo Dreyfus-Schmidt and Lewis Tunstall and Luigi Pagani and Moreira Machado and Pauline Bourigault and Ran Wang and Stanislas Polu and Thibaut Barroyer and Wen-Ding Li and Yazhe Niu and Yann Fleureau and Yangyang Hu and Zhouliang Yu and Zihan Wang and Zhilin Yang and Zhengying Liu and Jia Li},
      year={2025},
      eprint={2504.11354},
      archivePrefix={arXiv},
      primaryClass={cs.AI},
      url={https://arxiv.org/abs/2504.11354}, 
}

@article{LiquidTensor,
  author       = {Peter Scholze},
  title        = {Liquid Tensor Experiment},
  journal      = {Exp. Math.},
  volume       = {31},
  number       = {2},
  pages        = {349--354},
  year         = {2022},
  url          = {https://doi.org/10.1080/10586458.2021.1926016},
  doi          = {10.1080/10586458.2021.1926016},
  bibsource    = {dblp computer science bibliography, https://dblp.org}
}

@article{LiveBench,
  author       = {Colin White and
                  Samuel Dooley and
                  Manley Roberts and
                  Arka Pal and
                  Benjamin Feuer and
                  Siddhartha Jain and
                  Ravid Shwartz{-}Ziv and
                  Neel Jain and
                  Khalid Saifullah and
                  Siddartha Naidu and
                  Chinmay Hegde and
                  Yann LeCun and
                  Tom Goldstein and
                  Willie Neiswanger and
                  Micah Goldblum},
  title        = {LiveBench: {A} Challenging, Contamination-Free {LLM} Benchmark},
  journal      = {CoRR},
  volume       = {abs/2406.19314},
  year         = {2024},
  url          = {https://doi.org/10.48550/arXiv.2406.19314},
  doi          = {10.48550/ARXIV.2406.19314},
  eprinttype    = {arXiv},
  eprint       = {2406.19314},
  bibsource    = {dblp computer science bibliography, https://dblp.org}
}

@inproceedings{MixEval,
  author       = {Jinjie Ni and
                  Fuzhao Xue and
                  Xiang Yue and
                  Yuntian Deng and
                  Mahir Shah and
                  Kabir Jain and
                  Graham Neubig and
                  Yang You},
  editor       = {Amir Globersons and
                  Lester Mackey and
                  Danielle Belgrave and
                  Angela Fan and
                  Ulrich Paquet and
                  Jakub M. Tomczak and
                  Cheng Zhang},
  title        = {MixEval: Deriving Wisdom of the Crowd from {LLM} Benchmark Mixtures},
  booktitle    = {Advances in Neural Information Processing Systems 38: Annual Conference
                  on Neural Information Processing Systems 2024, NeurIPS 2024, Vancouver,
                  BC, Canada, December 10 - 15, 2024},
  year         = {2024},
  url          = {http://papers.nips.cc/paper\_files/paper/2024/hash/b1f34d7b4a03a3d80be8e72eb430dd81-Abstract-Conference.html},
  bibsource    = {dblp computer science bibliography, https://dblp.org}
}

@misc{TerryTao,
	author = {Terence Tao},
	title = {ams.org},
	howpublished = {\url{https://www.ams.org/notices/202501/rnoti-p6.pdf}},
	year = {2025},
	note = {[Accessed 10-05-2025]},
}

@article{GenLMforATP,
  author       = {Stanislas Polu and
                  Ilya Sutskever},
  title        = {Generative Language Modeling for Automated Theorem Proving},
  journal      = {CoRR},
  volume       = {abs/2009.03393},
  year         = {2020},
  url          = {https://arxiv.org/abs/2009.03393},
  eprinttype    = {arXiv},
  eprint       = {2009.03393},
  bibsource    = {dblp computer science bibliography, https://dblp.org}
}

@misc{TheoremProverJudge,
      title={Theorem Prover as a Judge for Synthetic Data Generation}, 
      author={Joshua Ong Jun Leang and Giwon Hong and Wenda Li and Shay B. Cohen},
      year={2025},
      eprint={2502.13137},
      archivePrefix={arXiv},
      primaryClass={cs.AI},
      url={https://arxiv.org/abs/2502.13137}, 
}

@misc{AntiLeakBench,
      title={AntiLeakBench: Preventing Data Contamination by Automatically Constructing Benchmarks with Updated Real-World Knowledge}, 
      author={Xiaobao Wu and Liangming Pan and Yuxi Xie and Ruiwen Zhou and Shuai Zhao and Yubo Ma and Mingzhe Du and Rui Mao and Anh Tuan Luu and William Yang Wang},
      year={2025},
      eprint={2412.13670},
      archivePrefix={arXiv},
      primaryClass={cs.CL},
      url={https://arxiv.org/abs/2412.13670}, 
}

@misc{PsychometricFTP,
      title={Psychometric-Based Evaluation for Theorem Proving with Large Language Models}, 
      author={Jianyu Zhang and Yongwang Zhao and Long Zhang and Jilin Hu and Xiaokun Luan and Zhiwei Xu and Feng Yang},
      year={2025},
      eprint={2502.00855},
      archivePrefix={arXiv},
      primaryClass={cs.AI},
      url={https://arxiv.org/abs/2502.00855}, 
}

@misc{Sonnet3.7,
  author = {{Anthropic}},
  title = {Claude 3.7 Sonnet System Card},
  howpublished = {\url{https://assets.anthropic.com/m/785e231869ea8b3b/original/claude-3-7-sonnet-system-card.pdf}},
  year = {2025},
  note = {Accessed: 2025-05-10}
}

@misc{DeepSeekProverV2,
      title={DeepSeek-Prover-V2: Advancing Formal Mathematical Reasoning via Reinforcement Learning for Subgoal Decomposition}, 
      author={Z. Z. Ren and Zhihong Shao and Junxiao Song and Huajian Xin and Haocheng Wang and Wanjia Zhao and Liyue Zhang and Zhe Fu and Qihao Zhu and Dejian Yang and Z. F. Wu and Zhibin Gou and Shirong Ma and Hongxuan Tang and Yuxuan Liu and Wenjun Gao and Daya Guo and Chong Ruan},
      year={2025},
      eprint={2504.21801},
      archivePrefix={arXiv},
      primaryClass={cs.CL},
      url={https://arxiv.org/abs/2504.21801}, 
}

@misc{Goedelprover,
          title={Goedel-Prover: A Frontier Model for Open-Source Automated Theorem Proving}, 
          author={Yong Lin and Shange Tang and Bohan Lyu and Jiayun Wu and Hongzhou Lin and Kaiyu Yang and Jia Li and Mengzhou Xia and Danqi Chen and Sanjeev Arora and Chi Jin},
          year={2025},
          eprint={2502.07640},
          archivePrefix={arXiv},
          primaryClass={cs.LG},
          url={https://arxiv.org/abs/2502.07640}, 
}

@inproceedings{miniF2F,
  author       = {Kunhao Zheng and
                  Jesse Michael Han and
                  Stanislas Polu},
  title        = {miniF2F: a cross-system benchmark for formal Olympiad-level mathematics},
  booktitle    = {The Tenth International Conference on Learning Representations, {ICLR}
                  2022, Virtual Event, April 25-29, 2022},
  publisher    = {OpenReview.net},
  year         = {2022},
  url          = {https://openreview.net/forum?id=9ZPegFuFTFv},
  bibsource    = {dblp computer science bibliography, https://dblp.org}
}

@inproceedings{INT,
    title={{\{}INT{\}}: An Inequality Benchmark for Evaluating Generalization in Theorem Proving},
    author={Yuhuai Wu and Albert Jiang and Jimmy Ba and Roger Baker Grosse},
    booktitle={International Conference on Learning Representations},
    year={2021},
    url={https://openreview.net/forum?id=O6LPudowNQm}
}

@inproceedings{MUSTARDSynthesis,
  author       = {Yinya Huang and
                  Xiaohan Lin and
                  Zhengying Liu and
                  Qingxing Cao and
                  Huajian Xin and
                  Haiming Wang and
                  Zhenguo Li and
                  Linqi Song and
                  Xiaodan Liang},
  title        = {{MUSTARD:} Mastering Uniform Synthesis of Theorem and Proof Data},
  booktitle    = {The Twelfth International Conference on Learning Representations,
                  {ICLR} 2024, Vienna, Austria, May 7-11, 2024},
  publisher    = {OpenReview.net},
  year         = {2024},
  url          = {https://openreview.net/forum?id=8xliOUg9EW},
  bibsource    = {dblp computer science bibliography, https://dblp.org}
}

@article{ProofNet,
  author       = {Zhangir Azerbayev and
                  Bartosz Piotrowski and
                  Hailey Schoelkopf and
                  Edward W. Ayers and
                  Dragomir Radev and
                  Jeremy Avigad},
  title        = {ProofNet: Autoformalizing and Formally Proving Undergraduate-Level
                  Mathematics},
  journal      = {CoRR},
  volume       = {abs/2302.12433},
  year         = {2023},
  url          = {https://doi.org/10.48550/arXiv.2302.12433},
  doi          = {10.48550/ARXIV.2302.12433},
  eprinttype    = {arXiv},
  eprint       = {2302.12433},
  bibsource    = {dblp computer science bibliography, https://dblp.org}
}

@inproceedings{LeanWorkbook,
  author       = {Huaiyuan Ying and
                  Zijian Wu and
                  Yihan Geng and
                  Jiayu Wang and
                  Dahua Lin and
                  Kai Chen},
  editor       = {Amir Globersons and
                  Lester Mackey and
                  Danielle Belgrave and
                  Angela Fan and
                  Ulrich Paquet and
                  Jakub M. Tomczak and
                  Cheng Zhang},
  title        = {Lean Workbook: {A} large-scale Lean problem set formalized from natural
                  language math problems},
  booktitle    = {Advances in Neural Information Processing Systems 38: Annual Conference
                  on Neural Information Processing Systems 2024, NeurIPS 2024, Vancouver,
                  BC, Canada, December 10 - 15, 2024},
  year         = {2024},
  url          = {http://papers.nips.cc/paper\_files/paper/2024/hash/bf236666a2cc5f3ae05d2e08485efc4c-Abstract-Datasets\_and\_Benchmarks\_Track.html},
  bibsource    = {dblp computer science bibliography, https://dblp.org}
}

@inproceedings{PutnamBench,
  author       = {George Tsoukalas and
                  Jasper Lee and
                  John Jennings and
                  Jimmy Xin and
                  Michelle Ding and
                  Michael Jennings and
                  Amitayush Thakur and
                  Swarat Chaudhuri},
  editor       = {Amir Globersons and
                  Lester Mackey and
                  Danielle Belgrave and
                  Angela Fan and
                  Ulrich Paquet and
                  Jakub M. Tomczak and
                  Cheng Zhang},
  title        = {PutnamBench: Evaluating Neural Theorem-Provers on the Putnam Mathematical
                  Competition},
  booktitle    = {Advances in Neural Information Processing Systems 38: Annual Conference
                  on Neural Information Processing Systems 2024, NeurIPS 2024, Vancouver,
                  BC, Canada, December 10 - 15, 2024},
  year         = {2024},
  url          = {http://papers.nips.cc/paper\_files/paper/2024/hash/1582eaf9e0cf349e1e5a6ee453100aa1-Abstract-Datasets\_and\_Benchmarks\_Track.html},
  bibsource    = {dblp computer science bibliography, https://dblp.org}
}

@misc{CombiBench,
      title={CombiBench: Benchmarking LLM Capability for Combinatorial Mathematics}, 
      author={Junqi Liu and Xiaohan Lin and Jonas Bayer and Yael Dillies and Weijie Jiang and Xiaodan Liang and Roman Soletskyi and Haiming Wang and Yunzhou Xie and Beibei Xiong and Zhengfeng Yang and Jujian Zhang and Lihong Zhi and Jia Li and Zhengying Liu},
      year={2025},
      eprint={2505.03171},
      archivePrefix={arXiv},
      primaryClass={cs.AI},
      url={https://arxiv.org/abs/2505.03171}, 
}

@article{GPQA,
  author       = {David Rein and
                  Betty Li Hou and
                  Asa Cooper Stickland and
                  Jackson Petty and
                  Richard Yuanzhe Pang and
                  Julien Dirani and
                  Julian Michael and
                  Samuel R. Bowman},
  title        = {{GPQA:} {A} Graduate-Level Google-Proof Q{\&}A Benchmark},
  journal      = {CoRR},
  volume       = {abs/2311.12022},
  year         = {2023},
  url          = {https://doi.org/10.48550/arXiv.2311.12022},
  doi          = {10.48550/ARXIV.2311.12022},
  eprinttype    = {arXiv},
  eprint       = {2311.12022},
  bibsource    = {dblp computer science bibliography, https://dblp.org}
}

@article{DeepSeekR1,
  author       = {DeepSeek{-}AI and
                  Daya Guo and
                  Dejian Yang and
                  Haowei Zhang and
                  Junxiao Song and
                  Ruoyu Zhang and
                  Runxin Xu and
                  Qihao Zhu and
                  Shirong Ma and
                  Peiyi Wang and
                  Xiao Bi and
                  Xiaokang Zhang and
                  Xingkai Yu and
                  Yu Wu and
                  Z. F. Wu and
                  Zhibin Gou and
                  Zhihong Shao and
                  Zhuoshu Li and
                  Ziyi Gao and
                  Aixin Liu and
                  Bing Xue and
                  Bingxuan Wang and
                  Bochao Wu and
                  Bei Feng and
                  Chengda Lu and
                  Chenggang Zhao and
                  Chengqi Deng and
                  Chenyu Zhang and
                  Chong Ruan and
                  Damai Dai and
                  Deli Chen and
                  Dongjie Ji and
                  Erhang Li and
                  Fangyun Lin and
                  Fucong Dai and
                  Fuli Luo and
                  Guangbo Hao and
                  Guanting Chen and
                  Guowei Li and
                  H. Zhang and
                  Han Bao and
                  Hanwei Xu and
                  Haocheng Wang and
                  Honghui Ding and
                  Huajian Xin and
                  Huazuo Gao and
                  Hui Qu and
                  Hui Li and
                  Jianzhong Guo and
                  Jiashi Li and
                  Jiawei Wang and
                  Jingchang Chen and
                  Jingyang Yuan and
                  Junjie Qiu and
                  Junlong Li and
                  J. L. Cai and
                  Jiaqi Ni and
                  Jian Liang and
                  Jin Chen and
                  Kai Dong and
                  Kai Hu and
                  Kaige Gao and
                  Kang Guan and
                  Kexin Huang and
                  Kuai Yu and
                  Lean Wang and
                  Lecong Zhang and
                  Liang Zhao and
                  Litong Wang and
                  Liyue Zhang and
                  Lei Xu and
                  Leyi Xia and
                  Mingchuan Zhang and
                  Minghua Zhang and
                  Minghui Tang and
                  Meng Li and
                  Miaojun Wang and
                  Mingming Li and
                  Ning Tian and
                  Panpan Huang and
                  Peng Zhang and
                  Qiancheng Wang and
                  Qinyu Chen and
                  Qiushi Du and
                  Ruiqi Ge and
                  Ruisong Zhang and
                  Ruizhe Pan and
                  Runji Wang and
                  R. J. Chen and
                  R. L. Jin and
                  Ruyi Chen and
                  Shanghao Lu and
                  Shangyan Zhou and
                  Shanhuang Chen and
                  Shengfeng Ye and
                  Shiyu Wang and
                  Shuiping Yu and
                  Shunfeng Zhou and
                  Shuting Pan and
                  S. S. Li},
  title        = {DeepSeek-R1: Incentivizing Reasoning Capability in LLMs via Reinforcement
                  Learning},
  journal      = {CoRR},
  volume       = {abs/2501.12948},
  year         = {2025},
  url          = {https://doi.org/10.48550/arXiv.2501.12948},
  doi          = {10.48550/ARXIV.2501.12948},
  eprinttype    = {arXiv},
  eprint       = {2501.12948},
  bibsource    = {dblp computer science bibliography, https://dblp.org}
}

@article{KiminaLeanServer,
  title={Kimina Lean Server: Technical Report},
  author={Santos, Marco Dos and Wang, Haiming and de Saxc{\'e}, Hugues and Wang, Ran and Baksys, Mantas and Unsal, Mert and Liu, Junqi and Liu, Zhengying and Li, Jia},
  journal={arXiv preprint arXiv:2504.21230},
  year={2025}
}

@inproceedings{MATH,
  author       = {Dan Hendrycks and
                  Collin Burns and
                  Saurav Kadavath and
                  Akul Arora and
                  Steven Basart and
                  Eric Tang and
                  Dawn Song and
                  Jacob Steinhardt},
  editor       = {Joaquin Vanschoren and
                  Sai{-}Kit Yeung},
  title        = {Measuring Mathematical Problem Solving With the {MATH} Dataset},
  booktitle    = {Proceedings of the Neural Information Processing Systems Track on
                  Datasets and Benchmarks 1, NeurIPS Datasets and Benchmarks 2021, December
                  2021, virtual},
  year         = {2021},
  url          = {https://datasets-benchmarks-proceedings.neurips.cc/paper/2021/hash/be83ab3ecd0db773eb2dc1b0a17836a1-Abstract-round2.html},
  bibsource    = {dblp computer science bibliography, https://dblp.org}
}

@misc{o3,
  title={OpenAI o3 and o4-mini System Card},
  author={OpenAI},
  year={2025},
  howpublished={\url{https://cdn.openai.com/pdf/2221c875-02dc-4789-800b-e7758f3722c1/o3-and-o4-mini-system-card.pdf}}
}

@misc{NPHardEval,
      title={NPHardEval: Dynamic Benchmark on Reasoning Ability of Large Language Models via Complexity Classes}, 
      author={Lizhou Fan and Wenyue Hua and Lingyao Li and Haoyang Ling and Yongfeng Zhang},
      year={2024},
      eprint={2312.14890},
      archivePrefix={arXiv},
      primaryClass={cs.AI},
      url={https://arxiv.org/abs/2312.14890}, 
}

@misc{Gemini2.5,
  author       = {{Google DeepMind}},
  title        = {Gemini 2.5: Our Most Intelligent AI Model},
  howpublished = {Google AI Blog, Mar 2025; \url{https://blog.google/technology/google-deepmind/gemini-model-thinking-updates-march-2025/}},
  year         = {2025}
}

@inproceedings{Lean4,
  author       = {Leonardo de Moura and
                  Sebastian Ullrich},
  editor       = {Andr{\'{e}} Platzer and
                  Geoff Sutcliffe},
  title        = {The Lean 4 Theorem Prover and Programming Language},
  booktitle    = {Automated Deduction - {CADE} 28 - 28th International Conference on
                  Automated Deduction, Virtual Event, July 12-15, 2021, Proceedings},
  series       = {Lecture Notes in Computer Science},
  volume       = {12699},
  pages        = {625--635},
  publisher    = {Springer},
  year         = {2021},
  url          = {https://doi.org/10.1007/978-3-030-79876-5\_37},
  doi          = {10.1007/978-3-030-79876-5\_37},
  bibsource    = {dblp computer science bibliography, https://dblp.org}
}

@book{Isabelle,
  author       = {Tobias Nipkow and
                  Lawrence C. Paulson and
                  Markus Wenzel},
  title        = {Isabelle/HOL - {A} Proof Assistant for Higher-Order Logic},
  series       = {Lecture Notes in Computer Science},
  volume       = {2283},
  publisher    = {Springer},
  year         = {2002},
  url          = {https://doi.org/10.1007/3-540-45949-9},
  doi          = {10.1007/3-540-45949-9},
  isbn         = {3-540-43376-7},
  bibsource    = {dblp computer science bibliography, https://dblp.org}
}

@misc{LeanRAG,
      title={Formal Language Knowledge Corpus for Retrieval Augmented Generation}, 
      author={Majd Zayyad and Yossi Adi},
      year={2024},
      eprint={2412.16689},
      archivePrefix={arXiv},
      primaryClass={cs.AI},
      url={https://arxiv.org/abs/2412.16689}, 
}

@article{CodeGeneration,
  author       = {William Murphy and
                  Nikolaus Holzer and
                  Feitong Qiao and
                  Leyi Cui and
                  Raven Rothkopf and
                  Nathan Koenig and
                  Mark Santolucito},
  title        = {Combining {LLM} Code Generation with Formal Specifications and Reactive
                  Program Synthesis},
  journal      = {CoRR},
  volume       = {abs/2410.19736},
  year         = {2024},
  url          = {https://doi.org/10.48550/arXiv.2410.19736},
  doi          = {10.48550/ARXIV.2410.19736},
  eprinttype    = {arXiv},
  eprint       = {2410.19736},
  bibsource    = {dblp computer science bibliography, https://dblp.org}
}

@misc{FormalAgent,
      title={Formal-LLM: Integrating Formal Language and Natural Language for Controllable LLM-based Agents}, 
      author={Zelong Li and Wenyue Hua and Hao Wang and He Zhu and Yongfeng Zhang},
      year={2024},
      eprint={2402.00798},
      archivePrefix={arXiv},
      primaryClass={cs.LG},
      url={https://arxiv.org/abs/2402.00798}, 
}

@inproceedings{AutoFormalizationLLM,
  author       = {Yuhuai Wu and
                  Albert Qiaochu Jiang and
                  Wenda Li and
                  Markus N. Rabe and
                  Charles Staats and
                  Mateja Jamnik and
                  Christian Szegedy},
  editor       = {Sanmi Koyejo and
                  S. Mohamed and
                  A. Agarwal and
                  Danielle Belgrave and
                  K. Cho and
                  A. Oh},
  title        = {Autoformalization with Large Language Models},
  booktitle    = {Advances in Neural Information Processing Systems 35: Annual Conference
                  on Neural Information Processing Systems 2022, NeurIPS 2022, New Orleans,
                  LA, USA, November 28 - December 9, 2022},
  year         = {2022},
  url          = {http://papers.nips.cc/paper\_files/paper/2022/hash/d0c6bc641a56bebee9d985b937307367-Abstract-Conference.html},
  bibsource    = {dblp computer science bibliography, https://dblp.org}
}
\bibliographystyle{icml2025}

\newpage
\appendix
\onecolumn
\section{Limitations}
While this work has been conducted according to best practices of previous work on formal theorem proving, there are several potential limitations: 

Firstly, our evaluation for proprietary OpenAI models are conducted with their default configuration, combined with our zero-temperature setting for open-sourced models to enhance reproducibility, all evaluated models may not have shared completely identical settings due to the undisclosed parameter settings of proprietary models.

Meanwhile, in order to strike a balance between budgetary restraints and performance representativeness, we only selected best-performing models on mainstream formal theorem proving benchmarks such as PutnamBench \cite{PutnamBench} and miniF2F \cite{miniF2F} for evaluation. There may be merit in exploring how other frontier proprietary models such as Gemini-2.5 \cite{Gemini2.5} and Claude-3.7-Sonnet \cite{Sonnet3.7} would perform on our dataset. Therefore, we plan to open-source our codebase and dataset to the community for researchers with ample resources to explore more models.


Lastly, while we leverage Lean compiler to automate proof validation, the proof is graded on a pass/fail basis as a whole. However, we note it's possible to implement a stepwise grading metric by dynamically masking each step in a whole proof and ask the model to fill in each step respectively, thereby gauging models' capability to implement key intermediate steps.

\section{Automated Proof Validation}\label{VerifyProof}
In our validation pipeline, we adopt the open-source Lean 4 server implementation from Kimina Lean Server \cite{KiminaLeanServer}. 
The Kimina Lean Server provides a Python interface allowing real-time feedback for generated proofs. With a Python function call \lstinline{verify}, the server receives a list of Lean proofs and returns validation results for each proof. The server itself handles multiple requests efficiently by spreading verification across multiple Lean REPL processes.

\section{Complete Example with Our Prompting Strategy}\label{prompt}
\begin{lstlisting}
{lean_code}
```
You can make your own auxiliary corollaries and theorems to support the proof, instead of only completing the part with the sorry. Please output the entire program and not just the last part. Please output only the program and add no other comment, such that your answer is a compilable lean code. Make sure to reason enough to make your code correct.
"""
\end{lstlisting}

 \paragraph{System Prompt} We used default system prompts as in the evaluated models' metadata.

\paragraph{A complete BB example.} \label{FullCase}
Below we demonstrate a complete record of evaluation, including the problem statement in Lean 4 and Markdown, our evaluation prompt and the model response:

\begin{lstlisting}
import Mathlib.Computability.TuringMachine

inductive Γ
  | zero
  | one
   deriving DecidableEq

instance : Inhabited Γ := ⟨ Γ.zero ⟩

structure Stmt where
  move : Turing.Dir
  write : Γ

def Machine := Fin 2 → Γ → Option (Fin 2 × Stmt)

structure Cfg where
  q : Fin 2
  Tape : Turing.Tape Γ

def init (l : List Γ) : Cfg := ⟨⟨0, by omega⟩, Turing.Tape.mk₁ l⟩

def step (M : Machine) : Cfg → Option Cfg :=
  fun ⟨q, T⟩ ↦ (M q T.head).map fun ⟨q', a⟩ ↦ ⟨q', (T.write a.write).move a.move⟩

def machine : Machine
| ⟨000, _⟩, Γ.zero  => some ⟨⟨000, by omega⟩, ⟨Turing.Dir.left, Γ.zero⟩⟩
| ⟨000, _⟩, Γ.one  => none
| ⟨001, _⟩, Γ.zero  => none
| ⟨001, _⟩, Γ.one  => some ⟨⟨000, by omega⟩, ⟨Turing.Dir.left, Γ.one⟩⟩
| ⟨_+2, _⟩, _ => by omega

def nth_cfg : (n : Nat) -> Option Cfg
| 0 => init []
| Nat.succ n => match (nth_cfg n) with
                | none => none
                | some cfg =>  step machine cfg

/-- 
Prove the following Turing Machine with the transition table never halts.
| State | Symbol | Next State | Move | Write |
|------|------|------|------|------|
| 000 | zero | 000 | left | zero |
| 000 | one | HALT | - | - |
| 001 | zero | HALT | - | - |
| 001 | one | 000 | left | one |
-/
theorem machine_never_halts : ∀ n, (nth_cfg n).isSome := by
  sorry
```

You can make your own auxiliary corollaries and theorems to support the proof, instead of only completing the part with the sorry. Please output the entire program and not just the last part. Please output only the program and add no other comment, such that your answer is a compilable lean code. Make sure to reason enough to make your code correct.
### Response:

```lean4
import Mathlib
import Aesop

set_option maxHeartbeats 0

open BigOperators Real Nat Topology Rat

/-- Given a sequence {aₙ} with a₁ = 1 and aₙ₊₁ = 2aₙ + 1,
    prove that aₙ = 2^n - 1 for all n ≥ 1 -/
theorem sequence_formula {a : ℕ → ℕ} (h1 : a 1 = 1)
    (h_rec : ∀ n : ℕ, n ≥ 1 → a (n + 1) = 2 * a n + 1) :
    ∀ n : ℕ, n ≥ 1 → a n = 2^n - 1 := by
  /-
  We need to prove that for a sequence \( \{a_n\} \) defined by \( a_1 = 1 \) and \( a_{n+1} = 2a_n + 1 \), the general term \( a_n \) is given by \( a_n = 2^n - 1 \) for all \( n \geq 1 \).
  1. **Base Case**: For \( n = 1 \), we have \( a_1 = 1 \). This matches \( 2^1 - 1 = 1 \), so the base case holds.
  2. **Inductive Step**: Assume \( a_n = 2^n - 1 \) holds for some \( n \geq 1 \). We need to show it holds for \( n + 1 \).
     - By the recurrence relation, \( a_{n+1} = 2a_n + 1 \).
     - Using the inductive hypothesis \( a_n = 2^n - 1 \), we get:
       \[
       a_{n+1} = 2(2^n - 1) + 1 = 2^{n+1} - 2 + 1 = 2^{n+1} - 1
       \]
     - Thus, \( a_{n+1} = 2^{n+1} - 1 \), completing the inductive step.
  By induction, \( a_n = 2^n - 1 \) holds for all \( n \geq 1 \).
  -/
  intro n hn
  induction' hn with n hn
  -- Base case: n = 1
  -- We know a 1 = 1, which matches 2^1 - 1 = 1
  simp_all
  -- Inductive step: Assume a n = 2^n - 1, prove a (n + 1) = 2^(n + 1) - 1
  cases n with
  | zero =>
    -- When n = 0, the induction hypothesis does not apply, so we directly use the initial condition
    simp_all
  | succ n =>
    -- For n = succ n, use the recurrence relation and the inductive hypothesis
    simp_all [Nat.succ_eq_add_one, Nat.pow_succ, Nat.mul_succ]
    -- Simplify the expression to match the form 2^(n + 1) - 1
    ring
    <;> omega
\end{lstlisting}


\paragraph{A complete step lemma challenge example.} \label{Step Lemma Full Case}
Below we demonstrate a complete record of evaluation, including the problem statement in Lean 4 and Markdown, our evaluation prompt and the model response:
\begin{lstlisting}
Theorem library:

```lean4
import Mathlib.Tactic.Lemma
import Mathlib.Tactic.NthRewrite
        
@[simp]
theorem bv32_and_not_self(x : BitVec 32) :
  x &&& ~~~x = 0 := by
  simp

@[simp]
theorem bv32_not_not(x : BitVec 32) :
  ~~~~~~x = x := by
  simp

@[simp]
theorem bv32_or_not_self(x : BitVec 32) :
  x ||| ~~~x = BitVec.allOnes 32 := by
  simp

@[simp]
theorem bv32_not_or_self(x : BitVec 32) :
  ~~~x ||| x = BitVec.allOnes 32 := by
  simp

@[simp]
theorem bv32_neg_mul (x y : BitVec 32) :
  -x * y = -(x * y) := by
  simp

theorem bv32_not_and (x y : BitVec 32) :
  ~~~(x &&& y) = ~~~x ||| ~~~y := by
  rw [BitVec.not_and]

theorem bv32_not_or (x y : BitVec 32) :
  ~~~(x ||| y) = ~~~x &&& ~~~y := by
  rw [BitVec.not_or]

theorem bv32_not_xor_eq_or (x y : (BitVec 32)) :
  ~~~(x ^^^ y) = (~~~x &&& ~~~y) ||| (x &&& y) := by
  ext i
  simp
  cases h₁ : x[i] <;> cases h₂ : y[i]
  simp
  simp
  simp
  simp

theorem bv32_xor_eq_or (x y : (BitVec 32)) :
  (x ^^^ y) = (~~~x &&& y) ||| (x &&& ~~~y) := by
  ext i
  simp
  cases h₁ : x[i] <;> cases h₂ : y[i]
  simp
  simp
  simp
  simp

theorem bv32_x_distr (x y: BitVec 32) :
  x = (x &&& y) ||| (x &&& ~~~y) := by
  ext i
  simp
  simp [← Bool.and_or_distrib_left]

theorem bv32_y_distr (x y: BitVec 32) :
  y = (x &&& y ||| ~~~x &&& y) := by
  ext i
  simp
  simp [← Bool.and_or_distrib_right]


theorem bv32_add_assoc (x y z : BitVec 32) :
  x + y + z = x + (y + z) := by
  rw [BitVec.add_assoc]

theorem bv32_add_comm(x y : BitVec 32) :
  x + y = y + x := by
  rw [BitVec.add_comm]

theorem bv32_add_neg_eq_sub {x y : BitVec 32} :
  x + -y = x - y := by
  rw [BitVec.add_neg_eq_sub]

theorem bv32_mul_comm (x y : BitVec 32) :
  x * y = y * x := by
  rw [BitVec.mul_comm]

theorem bv32_var_mul_comm (x y z: BitVec 32) : 
  (x &&& y) * z = z * (x &&& y) := by
  rw [BitVec.mul_comm]

theorem bv32_mul_add (x y z : BitVec 32) :
  x * (y + z) = x * y + x * z := by
  rw [BitVec.mul_add]

theorem bv32_neg_eq_mul (x : BitVec 32) :
  -x = x *  (BitVec.allOnes 32) := by
  rw [← BitVec.neg_one_eq_allOnes]
  rw [BitVec.mul_neg]
  rw [BitVec.mul_one]

theorem bv32_add_mul_one (x y : BitVec 32) :
  x + x * y = x * (1#32 + y) := by
  rw [BitVec.mul_add]
  rw [BitVec.mul_one]

/--
1: x &&& y
2: ~~~x &&& y
3: x &&& ~~~y
4: ~~~x &&& ~~~y
-/
theorem bv32_or_eq_add12 (x y : BitVec 32) :
  (x &&& y) ||| (~~~x &&& y) = (x &&& y) + (~~~x &&& y) := by
  apply Eq.symm
  apply BitVec.add_eq_or_of_and_eq_zero
  simp [← BitVec.and_assoc]
  simp [BitVec.and_comm _ (~~~x)]
  simp [← BitVec.and_assoc]

theorem bv32_or_eq_add13 (x y : BitVec 32) :
  (x &&& y) ||| (x &&& ~~~y) = (x &&& y) + (x &&& ~~~y) := by
  apply Eq.symm
  apply BitVec.add_eq_or_of_and_eq_zero
  simp [← BitVec.and_assoc]
  simp [BitVec.and_comm _ x]
  simp [BitVec.and_assoc]

theorem bv32_or_eq_add14 (x y : BitVec 32) :
  (x &&& y) ||| (~~~x &&& ~~~y) = (x &&& y) + (~~~x &&& ~~~y) := by
  apply Eq.symm 
  apply BitVec.add_eq_or_of_and_eq_zero
  simp [← BitVec.and_assoc]
  simp [BitVec.and_comm _ (~~~x)]
  simp [← BitVec.and_assoc]

theorem bv32_or_eq_add21 (x y : BitVec 32) :
  (~~~x &&& y) ||| (x &&& y) = (~~~x &&& y) + (x &&& y) := by
  apply Eq.symm
  apply BitVec.add_eq_or_of_and_eq_zero
  simp [← BitVec.and_assoc]
  simp [BitVec.and_comm _ x]
  simp [← BitVec.and_assoc]

theorem bv32_or_eq_add23 (x y : BitVec 32) :
  (~~~x &&& y) ||| (x &&& ~~~y) = (~~~x &&& y) + (x &&& ~~~y) := by
  apply Eq.symm
  apply BitVec.add_eq_or_of_and_eq_zero
  simp [← BitVec.and_assoc]
  simp [BitVec.and_comm _ x]
  simp [← BitVec.and_assoc]

theorem bv32_or_eq_add31 (x y : BitVec 32) :
  (x &&& ~~~y) ||| (x &&& y) = (x &&& ~~~y) + (x &&& y) := by
  apply Eq.symm
  apply BitVec.add_eq_or_of_and_eq_zero
  simp [← BitVec.and_assoc]
  simp [BitVec.and_comm _ x]
  simp [← BitVec.and_assoc]
  simp [BitVec.and_assoc]

theorem bv32_or_eq_add32 (x y : BitVec 32) :
  (x &&& ~~~y) ||| (~~~x &&& y) = (x &&& ~~~y) + (~~~x &&& y) := by
  apply Eq.symm
  apply BitVec.add_eq_or_of_and_eq_zero
  simp [← BitVec.and_assoc]
  simp [BitVec.and_comm _ (~~~x)]
  simp [← BitVec.and_assoc]

theorem bv32_or_eq_add41 (x y : BitVec 32) :
  (~~~x &&& ~~~y) ||| (x &&& y) = (~~~x &&& ~~~y) + (x &&& y) := by
  apply Eq.symm
  apply BitVec.add_eq_or_of_and_eq_zero
  simp [← BitVec.and_assoc]
  simp [BitVec.and_comm _ x]
  simp [← BitVec.and_assoc]

theorem bv32_or_eq_add_three (x y : BitVec 32) : 
  (x ||| y) = (x &&& ~~~y) + (x &&& y) + (~~~x &&& y) := by
  nth_rw 1 [bv32_y_distr x y]
  nth_rw 1 [bv32_x_distr x y]
  simp [← BitVec.or_assoc]
  simp [BitVec.or_comm _ (x &&& y)]
  simp [← BitVec.or_assoc]
  rw [BitVec.or_comm (x &&& y)]
  apply Eq.symm
  rw [BitVec.add_eq_or_of_and_eq_zero]
  rw [BitVec.add_eq_or_of_and_eq_zero]
  simp [← BitVec.and_assoc]
  simp [BitVec.and_comm _ x]
  simp [BitVec.and_assoc]
  rw [BitVec.add_comm]
  rw [← bv32_or_eq_add13]
  rw [← bv32_x_distr x y]
  simp [← BitVec.and_assoc]

theorem bv32_sum_all (x y : BitVec 32) :
  (~~~x &&& ~~~y) + (~~~x &&& y) + (x &&& y) + (x &&& ~~~y) = BitVec.allOnes 32 := by
  simp [BitVec.add_comm _ (~~~x &&& y)]
  simp [BitVec.add_comm _ (x &&& _)]
  simp [← BitVec.add_assoc]
  rw [BitVec.add_eq_or_of_and_eq_zero]
  rw [← bv32_or_eq_add_three x y]
  nth_rw 1 [bv32_x_distr x y]
  simp [BitVec.or_comm _ y]
  nth_rw 1 [bv32_y_distr x y]
  simp [← BitVec.or_assoc]
  simp [BitVec.or_comm _ (x &&& y)]
  simp [← BitVec.or_assoc]
  simp [BitVec.or_comm _ (~~~x &&& _)]
  simp [← BitVec.or_assoc]
  simp [BitVec.or_comm _ (~~~x &&& y)]
  rw [← bv32_x_distr (~~~x) y]
  rw [BitVec.or_assoc]
  rw [← bv32_x_distr x y]
  simp
  rw [← bv32_or_eq_add_three x y]
  simp [← BitVec.not_or]
  
theorem bv32_self_eq_neg_mul (x: BitVec 32):
  x = -x * (BitVec.allOnes 32) := by
  rw [BitVec.neg_mul]
  rw [BitVec.mul_comm]
  rw [← BitVec.neg_mul]
  simp [← BitVec.neg_one_eq_allOnes]
  
theorem bv32_not_self_and_not (x y : BitVec 32) :
  ~~~(x &&& ~~~x) = (~~~x &&& ~~~y) + (~~~x &&& y) + (x &&& y) + (x &&& ~~~y) := by
  rw [bv32_not_and]
  rw [BitVec.not_not]
  rw [bv32_not_or_self]
  rw [bv32_sum_all]
  
/-- 
Let x,y be 32-bit bit-vectors. Prove the equivalence of the following two expressions: $2\cdot \lnot (x\land \lnot y)-1\cdot (x\lor \lnot y)-7\cdot \lnot (x\oplus y)+11\cdot \lnot (x\land y)-5\cdot \lnot (x\lor y)-13\cdot \lnot (x\lor \lnot y)+6\cdot (x\land y)$, $10\cdot (x\land \lnot y)$
-/ 
theorem mba_challenge_02a2f35e (x y : BitVec 32) :  2#32 * ~~~(x &&& ~~~y) - 1#32 * (x ||| ~~~y) - 7#32 * ~~~(x ^^^ y) + 11#32 * ~~~(x &&& y) - 5#32 * ~~~(x ||| y) - 13#32 * ~~~(x ||| ~~~y) + 6#32 * (x &&& y) = 10#32 * (x &&& ~~~y) := by
  simp only [← bv32_add_neg_eq_sub] /- step 1 -/
  simp only [← bv32_neg_mul] /- step 2 -/
  simp only [bv32_not_and] /- step 3 -/
  simp only [bv32_not_or] /- step 4 -/
  sorry

lemma mba_challenge_02a2f35e_lhs_step_1 (x y : BitVec 32) : 2#32 * ~~~(x &&& ~~~y) - 1#32 * (x ||| ~~~y) - 7#32 * ~~~(x ^^^ y) + 11#32 * ~~~(x &&& y) - 5#32 * ~~~(x ||| y) - 13#32 * ~~~(x ||| ~~~y) + 6#32 * (x &&& y) = 2#32 * ~~~(x &&& ~~~y) + -(1#32 * (x ||| ~~~y)) + -(7#32 * ~~~(x ^^^ y)) + 11#32 * ~~~(x &&& y) + -(5#32 * ~~~(x ||| y)) + -(13#32 * ~~~(x ||| ~~~y)) + 6#32 * (x &&& y) := by
  simp only [← bv32_add_neg_eq_sub] 

lemma mba_challenge_02a2f35e_lhs_step_2 (x y : BitVec 32) : 2#32 * ~~~(x &&& ~~~y) + -(1#32 * (x ||| ~~~y)) + -(7#32 * ~~~(x ^^^ y)) + 11#32 * ~~~(x &&& y) + -(5#32 * ~~~(x ||| y)) + -(13#32 * ~~~(x ||| ~~~y)) + 6#32 * (x &&& y) = 2#32 * ~~~(x &&& ~~~y) + -1#32 * (x ||| ~~~y) + -7#32 * ~~~(x ^^^ y) + 11#32 * ~~~(x &&& y) + -5#32 * ~~~(x ||| y) + -13#32 * ~~~(x ||| ~~~y) + 6#32 * (x &&& y) := by
  simp only [← bv32_neg_mul] 

lemma mba_challenge_02a2f35e_lhs_step_3 (x y : BitVec 32) : 2#32 * ~~~(x &&& ~~~y) + -1#32 * (x ||| ~~~y) + -7#32 * ~~~(x ^^^ y) + 11#32 * ~~~(x &&& y) + -5#32 * ~~~(x ||| y) + -13#32 * ~~~(x ||| ~~~y) + 6#32 * (x &&& y) = 2#32 * (~~~x ||| ~~~~~~y) + -1#32 * (x ||| ~~~y) + -7#32 * ~~~(x ^^^ y) + 11#32 * (~~~x ||| ~~~y) + -5#32 * ~~~(x ||| y) + -13#32 * ~~~(x ||| ~~~y) + 6#32 * (x &&& y) := by
  simp only [bv32_not_and] 

lemma mba_challenge_02a2f35e_lhs_step_4 (x y : BitVec 32) : 2#32 * (~~~x ||| ~~~~~~y) + -1#32 * (x ||| ~~~y) + -7#32 * ~~~(x ^^^ y) + 11#32 * (~~~x ||| ~~~y) + -5#32 * ~~~(x ||| y) + -13#32 * ~~~(x ||| ~~~y) + 6#32 * (x &&& y) = 2#32 * (~~~x ||| ~~~~~~y) + -1#32 * (x ||| ~~~y) + -7#32 * ~~~(x ^^^ y) + 11#32 * (~~~x ||| ~~~y) + -5#32 * (~~~x &&& ~~~y) + -13#32 * (~~~x &&& ~~~~~~y) + 6#32 * (x &&& y) := by
  simp only [bv32_not_or] 

lemma mba_challenge_02a2f35e_lhs_step_5 (x y : BitVec 32) : 2#32 * (~~~x ||| ~~~~~~y) + -1#32 * (x ||| ~~~y) + -7#32 * ~~~(x ^^^ y) + 11#32 * (~~~x ||| ~~~y) + -5#32 * (~~~x &&& ~~~y) + -13#32 * (~~~x &&& ~~~~~~y) + 6#32 * (x &&& y) = 2#32 * (~~~x ||| ~~~~~~y) + -1#32 * (x ||| ~~~y) + -7#32 * (~~~x &&& ~~~y ||| x &&& y) + 11#32 * (~~~x ||| ~~~y) + -5#32 * (~~~x &&& ~~~y) + -13#32 * (~~~x &&& ~~~~~~y) + 6#32 * (x &&& y) := by
    simp only [<Theorem here>]
```

You are proving a single step of theorem 'mba_challenge_02a2f35e'. Based on the previous steps, select exactly one appropriate theorem from the Theorem library above and insert it in place of <Theorem here> to complete the proof of lemma 'mba_challenge_02a2f35e_lhs_step_5'. Do not modify any code from previous theorems or lemmas. You are also not allowed to add any other tactics to the proof body.
### Response:
You need the lemma that rewrites ~~~(x ^^^ y) into (~~~x &&& ~~~y) ||| (x &&& y). That is
bv32_not_xor_eq_or
So your step 5 becomes:
lemma mba_challenge_02a2f35e_lhs_step_5 (x y : BitVec 32) :
  2#32 * (~~~x ||| ~~~~~~y) + -1#32 * (x ||| ~~~y) + -7#32 * ~~~(x ^^^ y) + 11#32 * (~~~x ||| ~~~y) + -5#32 * (~~~x &&& ~~~y) + -13#32 * (~~~x &&& ~~~~~~y) + 6#32 * (x &&& y) = 2#32 * (~~~x ||| ~~~~~~y) + -1#32 * (x ||| ~~~y) + -7#32 * (~~~x &&& ~~~y ||| x &&& y) + 11#32 * (~~~x ||| ~~~y) + -5#32 * (~~~x &&& ~~~y) + -13#32 * (~~~x &&& ~~~~~~y) + 6#32 * (x &&& y) := by
  simp only [bv32_not_xor_eq_or]
This completes the requested rewrite.
\end{lstlisting}

\section{Template-Based Formal-Informal Translation}\label{Translation}
We use the following Python program to
produce problem descriptions in rigorously aligned Lean-Markdown pair.
\begin{lstlisting}
import srsly
from tqdm import tqdm

output_dir = "busy_beaver/prompts_to_run"

tables = list(srsly.read_jsonl("busy_beaver/bb_with_table.jsonl"))

template = """Complete the following Lean 4 code:

```lean4
{lean_code}
```

You can make your own auxiliary corollaries and theorems to support the proof, instead of only completing the part with the sorry. Please output the entire program and not just the last part. Please output only the program and add no other comment, such that your answer is a compilable lean code. Make sure to reason enough to make your code correct.
"""

informal_tmplate = """
/-- 
{informal}
-/
"""

for problem in tables:
    lean_path = problem["file_name"]
    file_name = lean_path.split("/")[-1]
    out_path = f"{output_dir}/{file_name.removesuffix('.lean')}.txt"
    is_halting = True
    if "nonhalting" in file_name:
        is_halting = False
    informal_head = "Prove the following Turing Machine with the transition table " + (
        "halts." if is_halting else "never halts."
    )
    # print(out_path)
    lean_code = None
    with open(lean_path, "r") as f:
        lean_code = f.read().removeprefix("\n")
    assert lean_code != None
    informal = "/-- \n" + informal_head + "\n" + problem["table"] + "-/\n"
    with open(out_path, "w") as f:
        f.write(
            template.format(
                lean_code=lean_code.replace(
                    "theorem machine_", informal + "theorem machine_"
                ).replace("\n\ndef nth_cfg", "\ndef nth_cfg")
                .replace("\n\n/--", "\n/--")
                .replace("sorry\n", "sorry")
            )
        )
        pass
\end{lstlisting}

\section{More Success and Failure Cases}
\label{AdditionalCase}
We demonstrate additional full record of evaluation, including problem statements in their Lean-Markdown pair and model responses for both successful and failed evaluation cases. The comprehensive record of model responses for cases can be found in the results folder of our dataset.
\subsection{Success Case}
\subsubsection{Deepseek-prover-v2-671B/bb-4state-case107-nonhalting} 
This is a trivial non-halting case, meaning there is no "HALT" entry in the transition table.  It can therefore be resolved via a single structural induction. Although the proof completes, warning messages appear, such as "aesop: failed to prove the goal after exhaustive search", "(try aesop) tactic does nothing", and "this tactic is never executed". These suggest the model does not fully reason about the proof and instead relies on automated tactics, hoping some may resolve the goal.
\begin{lstlisting}
import Mathlib.Computability.TuringMachine

inductive Γ
  | zero
  | one
   deriving DecidableEq

instance : Inhabited Γ := ⟨ Γ.zero ⟩

structure Stmt where
  move : Turing.Dir
  write : Γ

def Machine := Fin 4 → Γ → Option (Fin 4 × Stmt)

structure Cfg where
  q : Fin 4
  Tape : Turing.Tape Γ

def init (l : List Γ) : Cfg := ⟨⟨0, by omega⟩, Turing.Tape.mk₁ l⟩

def step (M : Machine) : Cfg → Option Cfg :=
  fun ⟨q, T⟩ ↦ (M q T.head).map fun ⟨q', a⟩ ↦ ⟨q', (T.write a.write).move a.move⟩

def machine : Machine
| ⟨000, _⟩, Γ.zero  => some ⟨⟨001, by omega⟩, ⟨Turing.Dir.left, Γ.one⟩⟩
| ⟨000, _⟩, Γ.one  => some ⟨⟨001, by omega⟩, ⟨Turing.Dir.right, Γ.one⟩⟩
| ⟨001, _⟩, Γ.zero  => some ⟨⟨000, by omega⟩, ⟨Turing.Dir.right, Γ.one⟩⟩
| ⟨001, _⟩, Γ.one  => some ⟨⟨003, by omega⟩, ⟨Turing.Dir.left, Γ.one⟩⟩
| ⟨002, _⟩, Γ.zero  => some ⟨⟨001, by omega⟩, ⟨Turing.Dir.right, Γ.zero⟩⟩
| ⟨002, _⟩, Γ.one  => some ⟨⟨000, by omega⟩, ⟨Turing.Dir.right, Γ.zero⟩⟩
| ⟨003, _⟩, Γ.zero  => some ⟨⟨000, by omega⟩, ⟨Turing.Dir.left, Γ.zero⟩⟩
| ⟨003, _⟩, Γ.one  => some ⟨⟨001, by omega⟩, ⟨Turing.Dir.left, Γ.zero⟩⟩
| ⟨_+4, _⟩, _ => by omega

def nth_cfg : (n : Nat) -> Option Cfg
| 0 => init []
| Nat.succ n => match (nth_cfg n) with
                | none => none
                | some cfg =>  step machine cfg

/-- 
Prove the following Turing Machine with the transition table never halts.
| State | Symbol | Next State | Move | Write |
|-----|-----|-----|-----|-----|
| 000 | zero | 001 | left | one |
| 000 | one | 001 | right | one |
| 001 | zero | 000 | right | one |
| 001 | one | 003 | left | one |
| 002 | zero | 001 | right | zero |
| 002 | one | 000 | right | zero |
| 003 | zero | 000 | left | zero |
| 003 | one | 001 | left | zero |
-/
theorem machine_never_halts : ∀ n, (nth_cfg n).isSome := by
  intro n
  have h : ∀ n, (nth_cfg n).isSome := by
    intro n
    induction n with
    | zero => simp [nth_cfg, init]
    | succ n ih =>
      cases h₁ : nth_cfg n with
      | none => simp_all [nth_cfg, step, machine, Cfg, Stmt, Γ, Fin.forall_fin_succ, Fin.forall_fin_succ']
      | some cfg =>
        simp_all [nth_cfg, step, machine, Cfg, Stmt, Γ, Fin.forall_fin_succ, Fin.forall_fin_succ']
        <;>
        (try cases cfg <;> simp_all [nth_cfg, step, machine, Cfg, Stmt, Γ, Fin.forall_fin_succ, Fin.forall_fin_succ'])
        <;>
        (try aesop)
        <;>
        (try omega)
        <;>
        (try aesop)
  exact h n
\end{lstlisting}

\subsection{Failure Case}
\subsubsection{Deepseek-prover-v2-671B/bb-2state-case102-nonhalting}  
This is a case where the Turing machine’s state remains fixed at 000. A correct proof requires showing that the machine consistently executes the first rule of the transition table. The model attempts a single structural induction; however, in the inductive case $succ \ n \ ih$, it leaves two unsolved goals $False$. These cannot be resolved via the induction hypothesis, suggesting that the model fails to generalize inductive reasoning in this setting.
\begin{lstlisting}
import Mathlib.Computability.TuringMachine

inductive Γ
  | zero
  | one
   deriving DecidableEq

instance : Inhabited Γ := ⟨ Γ.zero ⟩

structure Stmt where
  move : Turing.Dir
  write : Γ

def Machine := Fin 2 → Γ → Option (Fin 2 × Stmt)

structure Cfg where
  q : Fin 2
  Tape : Turing.Tape Γ

def init (l : List Γ) : Cfg := ⟨⟨0, by omega⟩, Turing.Tape.mk₁ l⟩

def step (M : Machine) : Cfg → Option Cfg :=
  fun ⟨q, T⟩ ↦ (M q T.head).map fun ⟨q', a⟩ ↦ ⟨q', (T.write a.write).move a.move⟩

def machine : Machine
| ⟨000, _⟩, Γ.zero  => some ⟨⟨000, by omega⟩, ⟨Turing.Dir.left, Γ.zero⟩⟩
| ⟨000, _⟩, Γ.one  => none
| ⟨001, _⟩, Γ.zero  => none
| ⟨001, _⟩, Γ.one  => some ⟨⟨000, by omega⟩, ⟨Turing.Dir.left, Γ.one⟩⟩
| ⟨_+2, _⟩, _ => by omega

def nth_cfg : (n : Nat) -> Option Cfg
| 0 => init []
| Nat.succ n => match (nth_cfg n) with
                | none => none
                | some cfg =>  step machine cfg

/-- 
Prove the following Turing Machine with the transition table never halts.
| State | Symbol | Next State | Move | Write |
|------|------|------|------|------|
| 000 | zero | 000 | left | zero |
| 000 | one | HALT | - | - |
| 001 | zero | HALT | - | - |
| 001 | one | 000 | left | one |
-/
theorem machine_never_halts : ∀ n, (nth_cfg n).isSome := by
  intro n
  have h : ∀ n, (nth_cfg n).isSome := by
    intro n
    induction n with
    | zero =>
      simp [nth_cfg, init]
    | succ n ih =>
      simp_all [nth_cfg, step, machine, Cfg.q, Cfg.Tape, Turing.Tape.move, Option.isSome]
      <;> aesop
  exact h n
\end{lstlisting}

\subsubsection{Goedel-prover/bb-2state-case102-nonhalting}  
The model produced proof of a completely unrelated theorem about sequences. This seems to be a hallucination, where the model matched a common pattern like $theorem ... by\ sorry$ and filled in a generic proof without understanding the actual task. This shows a key limitation of small provers: Without symbolic grounding or verification, they can only match surface-level patterns rather than truly reasoning.
\begin{lstlisting}
import Mathlib
import Aesop

set_option maxHeartbeats 0

open BigOperators Real Nat Topology Rat

/-- Given a sequence {aₙ} with a₁ = 1 and aₙ₊₁ = 2aₙ + 1,
    prove that aₙ = 2^n - 1 for all n ≥ 1 -/
theorem sequence_formula {a : ℕ → ℕ} (h1 : a 1 = 1)
    (h_rec : ∀ n : ℕ, n ≥ 1 → a (n + 1) = 2 * a n + 1) :
    ∀ n : ℕ, n ≥ 1 → a n = 2^n - 1 := by
  /-
  We need to prove that for a sequence \( \{a_n\} \) defined by \( a_1 = 1 \) and \( a_{n+1} = 2a_n + 1 \), the general term \( a_n \) is given by \( a_n = 2^n - 1 \) for all \( n \geq 1 \).
  1. **Base Case**: For \( n = 1 \), we have \( a_1 = 1 \). This matches \( 2^1 - 1 = 1 \), so the base case holds.
  2. **Inductive Step**: Assume \( a_n = 2^n - 1 \) holds for some \( n \geq 1 \). We need to show it holds for \( n + 1 \).
     - By the recurrence relation, \( a_{n+1} = 2a_n + 1 \).
     - Using the inductive hypothesis \( a_n = 2^n - 1 \), we get:
       \[
       a_{n+1} = 2(2^n - 1) + 1 = 2^{n+1} - 2 + 1 = 2^{n+1} - 1
       \]
     - Thus, \( a_{n+1} = 2^{n+1} - 1 \), completing the inductive step.
  By induction, \( a_n = 2^n - 1 \) holds for all \( n \geq 1 \).
  -/
  intro n hn
  induction' hn with n hn
  -- Base case: n = 1
  -- We know a 1 = 1, which matches 2^1 - 1 = 1
  simp_all
  -- Inductive step: Assume a n = 2^n - 1, prove a (n + 1) = 2^(n + 1) - 1
  cases n with
  | zero =>
    -- When n = 0, the induction hypothesis does not apply, so we directly use the initial condition
    simp_all
  | succ n =>
    -- For n = succ n, use the recurrence relation and the inductive hypothesis
    simp_all [Nat.succ_eq_add_one, Nat.pow_succ, Nat.mul_succ]
    -- Simplify the expression to match the form 2^(n + 1) - 1
    ring
    <;> omega
\end{lstlisting}

\subsubsection{Kimini-Prover-7B/bb-4state-case122-nonhalting}  
The model just outputs $sorry$, which means it didn’t complete the proof. This might be because it couldn’t figure out how to reason about the problem, or because it has seen $sorry$ used as a placeholder in Lean and assumed that was enough to count as a valid answer.
\begin{lstlisting}
import Mathlib.Computability.TuringMachine

inductive Γ
  | zero
  | one
   deriving DecidableEq

instance : Inhabited Γ := ⟨ Γ.zero ⟩

structure Statement where
  move : Turing Dir
  write : Γ

def Machine := Fin 4 → Γ → Option (Fin 4 × Statement)

structure Cfg where
  q : Fin 4
  Tape : Turing Tape Γ

def init (l : List Γ) : Cfg := ⟨⟨0, by omega⟩, ⟨Turing Dir.right, Γ.zero⟩⟩

def step (M : Machine) : Cfg → Option Cfg :=
  fun ⟨q, T⟩ ↦ (M q T.head).map fun ⟨q', a⟩ ↦ ⟨q', (T.map a).move a⟩

def machine : Machine
| ⟨000, _⟩, Γ.zero  => some ⟨⟨001, by omega⟩, ⟨Turing Dir.right, Γ.zero⟩⟩
| ⟨000, _⟩, Γ.one  => some ⟨⟨003, by omega⟩, ⟨Turing Dir.left, Γ.zero⟩⟩
| ⟨001, _⟩, Γ.zero  => some ⟨⟨003, by omega⟩, ⟨Turing Dir.left, Γ.one⟩⟩
| ⟨001, _⟩, Γ.one  => some ⟨⟨002, by omega⟩, ⟨Turing Dir.right, Γ.zero⟩⟩
| ⟨002, _⟩, Γ.zero  => some ⟨⟨003, by omega⟩, ⟨Turing Dir.left, Γ.zero⟩⟩
| ⟨002, _⟩, Γ.one  => some ⟨⟨001, by omega⟩, ⟨Turing Dir.left, Γ.one⟩⟩
| ⟨003, _⟩, Γ.zero  => some ⟨⟨001, by omega⟩, ⟨Turing Dir.left, Γ.one⟩⟩
| ⟨003, _⟩, Γ.one  => some ⟨⟨000, by omega⟩, ⟨Turing Dir.right, Γ.zero⟩⟩
| ⟨_+4, _⟩, _ => by omega

def nth configurations (n : Nat) : Option Cfg :=
  | 0 => init []
  | Nat.succ n => match (nth configurations n) with
                | none => none
                | some cfg =>  step machine cfg

/-- 
Prove the following Turing Machine with the transition table never halts.
| State | Symbol | Next State | Move | Write |
|-------|--------|------------|------|
| 000 | zero | 001 | right | zero |
| 000 | one | 003 | left | zero |
| 001 | zero | 003 | left | one |
| 001 | one | 002 | right | zero |
| 002 | zero | 003 | left | zero |
| 002 | one | 001 | left | one |
| 003 | zero | 001 | left | one |
| 003 | one | 000 | right | zero |
-/
theorem machine_never_halts : ∀ n, (nth configurations n).isSome := by
  sorry
\end{lstlisting}

\section{Customized Lemma Library.} \label{custom lemma library}

\begin{lstlisting}
import Mathlib.Tactic.NthRewrite
        
@[simp]
theorem bv32_and_not_self(x : BitVec 32) :
  x &&& ~~~x = 0 := by
  simp

@[simp]
theorem bv32_not_not(x : BitVec 32) :
  ~~~~~~x = x := by
  simp

@[simp]
theorem bv32_or_not_self(x : BitVec 32) :
  x ||| ~~~x = BitVec.allOnes 32 := by
  simp

@[simp]
theorem bv32_not_or_self(x : BitVec 32) :
  ~~~x ||| x = BitVec.allOnes 32 := by
  simp

@[simp]
theorem bv32_neg_mul (x y : BitVec 32) :
  -x * y = -(x * y) := by
  simp

theorem bv32_not_and (x y : BitVec 32) :
  ~~~(x &&& y) = ~~~x ||| ~~~y := by
  rw [BitVec.not_and]

theorem bv32_not_or (x y : BitVec 32) :
  ~~~(x ||| y) = ~~~x &&& ~~~y := by
  rw [BitVec.not_or]

theorem bv32_not_xor_eq_or (x y : (BitVec 32)) :
  ~~~(x ^^^ y) = (~~~x &&& ~~~y) ||| (x &&& y) := by
  ext i
  simp
  cases h₁ : x[i] <;> cases h₂ : y[i]
  simp
  simp
  simp
  simp

theorem bv32_xor_eq_or (x y : (BitVec 32)) :
  (x ^^^ y) = (~~~x &&& y) ||| (x &&& ~~~y) := by
  ext i
  simp
  cases h₁ : x[i] <;> cases h₂ : y[i]
  simp
  simp
  simp
  simp

theorem bv32_x_distr (x y: BitVec 32) :
  x = (x &&& y) ||| (x &&& ~~~y) := by
  ext i
  simp
  simp [← Bool.and_or_distrib_left]

theorem bv32_y_distr (x y: BitVec 32) :
  y = (x &&& y ||| ~~~x &&& y) := by
  ext i
  simp
  simp [← Bool.and_or_distrib_right]


theorem bv32_add_assoc (x y z : BitVec 32) :
  x + y + z = x + (y + z) := by
  rw [BitVec.add_assoc]

theorem bv32_add_comm(x y : BitVec 32) :
  x + y = y + x := by
  rw [BitVec.add_comm]

theorem bv32_add_neg_eq_sub {x y : BitVec 32} :
  x + -y = x - y := by
  rw [BitVec.add_neg_eq_sub]

theorem bv32_mul_comm (x y : BitVec 32) :
  x * y = y * x := by
  rw [BitVec.mul_comm]

theorem bv32_var_mul_comm (x y z: BitVec 32) : 
  (x &&& y) * z = z * (x &&& y) := by
  rw [BitVec.mul_comm]

theorem bv32_mul_add (x y z : BitVec 32) :
  x * (y + z) = x * y + x * z := by
  rw [BitVec.mul_add]

theorem bv32_neg_eq_mul (x : BitVec 32) :
  -x = x *  (BitVec.allOnes 32) := by
  rw [← BitVec.neg_one_eq_allOnes]
  rw [BitVec.mul_neg]
  rw [BitVec.mul_one]

theorem bv32_add_mul_one (x y : BitVec 32) :
  x + x * y = x * (1#32 + y) := by
  rw [BitVec.mul_add]
  rw [BitVec.mul_one]

/--
1: x &&& y
2: ~~~x &&& y
3: x &&& ~~~y
4: ~~~x &&& ~~~y
-/
theorem bv32_or_eq_add12 (x y : BitVec 32) :
  (x &&& y) ||| (~~~x &&& y) = (x &&& y) + (~~~x &&& y) := by
  apply Eq.symm
  apply BitVec.add_eq_or_of_and_eq_zero
  simp [← BitVec.and_assoc]
  simp [BitVec.and_comm _ (~~~x)]
  simp [← BitVec.and_assoc]

theorem bv32_or_eq_add13 (x y : BitVec 32) :
  (x &&& y) ||| (x &&& ~~~y) = (x &&& y) + (x &&& ~~~y) := by
  apply Eq.symm
  apply BitVec.add_eq_or_of_and_eq_zero
  simp [← BitVec.and_assoc]
  simp [BitVec.and_comm _ x]
  simp [BitVec.and_assoc]

theorem bv32_or_eq_add14 (x y : BitVec 32) :
  (x &&& y) ||| (~~~x &&& ~~~y) = (x &&& y) + (~~~x &&& ~~~y) := by
  apply Eq.symm
  apply BitVec.add_eq_or_of_and_eq_zero
  simp [← BitVec.and_assoc]
  simp [BitVec.and_comm _ (~~~x)]
  simp [← BitVec.and_assoc]

theorem bv32_or_eq_add21 (x y : BitVec 32) :
  (~~~x &&& y) ||| (x &&& y) = (~~~x &&& y) + (x &&& y) := by
  apply Eq.symm
  apply BitVec.add_eq_or_of_and_eq_zero
  simp [← BitVec.and_assoc]
  simp [BitVec.and_comm _ x]
  simp [← BitVec.and_assoc]

theorem bv32_or_eq_add23 (x y : BitVec 32) :
  (~~~x &&& y) ||| (x &&& ~~~y) = (~~~x &&& y) + (x &&& ~~~y) := by
  apply Eq.symm
  apply BitVec.add_eq_or_of_and_eq_zero
  simp [← BitVec.and_assoc]
  simp [BitVec.and_comm _ x]
  simp [← BitVec.and_assoc]

theorem bv32_or_eq_add31 (x y : BitVec 32) :
  (x &&& ~~~y) ||| (x &&& y) = (x &&& ~~~y) + (x &&& y) := by
  apply Eq.symm
  apply BitVec.add_eq_or_of_and_eq_zero
  simp [← BitVec.and_assoc]
  simp [BitVec.and_comm _ x]
  simp [← BitVec.and_assoc]
  simp [BitVec.and_assoc]

theorem bv32_or_eq_add32 (x y : BitVec 32) :
  (x &&& ~~~y) ||| (~~~x &&& y) = (x &&& ~~~y) + (~~~x &&& y) := by
  apply Eq.symm
  apply BitVec.add_eq_or_of_and_eq_zero
  simp [← BitVec.and_assoc]
  simp [BitVec.and_comm _ (~~~x)]
  simp [← BitVec.and_assoc]

theorem bv32_or_eq_add41 (x y : BitVec 32) :
  (~~~x &&& ~~~y) ||| (x &&& y) = (~~~x &&& ~~~y) + (x &&& y) := by
  apply Eq.symm
  apply BitVec.add_eq_or_of_and_eq_zero
  simp [← BitVec.and_assoc]
  simp [BitVec.and_comm _ x]
  simp [← BitVec.and_assoc]

theorem bv32_or_eq_add_three (x y : BitVec 32) : 
  (x ||| y) = (x &&& ~~~y) + (x &&& y) + (~~~x &&& y) := by
  nth_rw 1 [bv32_y_distr x y]
  nth_rw 1 [bv32_x_distr x y]
  simp [← BitVec.or_assoc]
  simp [BitVec.or_comm _ (x &&& y)]
  simp [← BitVec.or_assoc]
  rw [BitVec.or_comm (x &&& y)]
  apply Eq.symm
  rw [BitVec.add_eq_or_of_and_eq_zero]
  rw [BitVec.add_eq_or_of_and_eq_zero]
  simp [← BitVec.and_assoc]
  simp [BitVec.and_comm _ x]
  simp [BitVec.and_assoc]
  rw [BitVec.add_comm]
  rw [← bv32_or_eq_add13]
  rw [← bv32_x_distr x y]
  simp [← BitVec.and_assoc]

theorem bv32_sum_all (x y : BitVec 32) :
  (~~~x &&& ~~~y) + (~~~x &&& y) + (x &&& y) + (x &&& ~~~y) = BitVec.allOnes 32 := by
  simp [BitVec.add_comm _ (~~~x &&& y)]
  simp [BitVec.add_comm _ (x &&& _)]
  simp [← BitVec.add_assoc]
  rw [BitVec.add_eq_or_of_and_eq_zero]
  rw [← bv32_or_eq_add_three x y]
  nth_rw 1 [bv32_x_distr x y]
  simp [BitVec.or_comm _ y]
  nth_rw 1 [bv32_y_distr x y]
  simp [← BitVec.or_assoc]
  simp [BitVec.or_comm _ (x &&& y)]
  simp [← BitVec.or_assoc]
  simp [BitVec.or_comm _ (~~~x &&& _)]
  simp [← BitVec.or_assoc]
  simp [BitVec.or_comm _ (~~~x &&& y)]
  rw [← bv32_x_distr (~~~x) y]
  rw [BitVec.or_assoc]
  rw [← bv32_x_distr x y]
  simp
  rw [← bv32_or_eq_add_three x y]
  simp [← BitVec.not_or]
  
theorem bv32_self_eq_neg_mul (x: BitVec 32):
  x = -x * (BitVec.allOnes 32) := by
  rw [BitVec.neg_mul]
  rw [BitVec.mul_comm]
  rw [← BitVec.neg_mul]
  simp [← BitVec.neg_one_eq_allOnes]
  
theorem bv32_not_self_and_not (x y : BitVec 32) :
  ~~~(x &&& ~~~x) = (~~~x &&& ~~~y) + (~~~x &&& y) + (x &&& y) + (x &&& ~~~y) := by
  rw [bv32_not_and]
  rw [BitVec.not_not]
  rw [bv32_not_or_self]
  rw [bv32_sum_all]
\end{lstlisting}\label{sec:appendix}

\end{CJK}
\end{document}